\documentclass[letterpaper,12pt]{article}
\usepackage[english]{babel}
\usepackage{blindtext}
\usepackage[utf8]{inputenc}
\usepackage[labelformat=empty]{caption}
\usepackage{textcomp}
\usepackage{comment}
\usepackage{color, soul}
\usepackage{courier}
\usepackage[super]{cite}
\usepackage[title]{appendix}
\usepackage{enumitem}
\usepackage{booktabs}
\usepackage{makecell}
\usepackage{tabularx}
\usepackage{setspace}

\usepackage{amssymb}

\usepackage{dashrule}

\usepackage{achemso}

\usepackage{verbatim}

\usepackage{listings}
\usepackage{color, colortbl}
\usepackage[svgnames]{xcolor}

\usepackage{longtable}
\usepackage{tabu}
\usepackage{siunitx} 
\usepackage{pdflscape}

\input{listings-stata.tex}

\definecolor{codegreen}{rgb}{0,0.6,0}
\definecolor{codegray}{rgb}{0.5,0.5,0.5}
\definecolor{codepurple}{rgb}{0.58,0,0.82}
\definecolor{backcolour}{rgb}{0.95,0.95,0.92}
\definecolor{lightgray}{RGB}{211,211,211}
\definecolor{lgray2}{RGB}{182,182,185}

\lstdefinestyle{mystyle}{
    backgroundcolor=\color{backcolour},   
    commentstyle=\color{codegreen},
    keywordstyle=\color{magenta},
    numberstyle=\tiny\color{codegray},
    stringstyle=\color{codepurple},
    basicstyle=\ttfamily\footnotesize,
    breakatwhitespace=false,         
    breaklines=true,                 
    captionpos=b,                    
    keepspaces=true,                 
    numbers=left,                    
    numbersep=5pt,                  
    showspaces=false,                
    showstringspaces=false,
    showtabs=false,                  
    tabsize=2
}

\usepackage{graphicx}
\graphicspath{ {./images/} }

\usepackage{amsmath}
\usepackage[fleqn]{mathtools}

\usepackage{adjustbox}
\usepackage{afterpage}
\usepackage{tikz}
\usetikzlibrary{positioning, calc, shapes.geometric, shapes.multipart, shapes, arrows.meta, arrows, decorations.markings, external, trees}

\tikzstyle{Arrow} = [thick, decoration={markings,mark=at position 1 with {\arrow[thick]{latex}}},shorten >= 3pt, preaction = {decorate}]
\tikzstyle{Arrow2} = [thick, decoration={markings,mark=at position 0.9999 with {\arrow[thick]{latex}}},shorten >= 3pt, preaction = {decorate}]
\tikzstyle{Arrow3} = [thick, decoration={markings,mark=at position 0.9995 with {\arrow[thick]{latex}}},shorten >= 3pt, preaction = {decorate}]

\usepackage{blindtext}

\usepackage [autostyle, english = american]{csquotes}
\MakeOuterQuote{"}

\title{Assessing Outcome-to-Outcome Interference in Sibling Fixed Effects Models}
\author{David C. Mallinson \\
	Population Research Center, University of Texas-Austin}
\date{%
	\normalsize{\today}
	}

\begin{document}

\pagenumbering{gobble}

\maketitle

\small{}
\begin{center}
\textbf{Abstract}
\end{center}

\noindent Sibling fixed effects (FE) models are useful for estimating causal treatment effects while offsetting unobserved sibling-invariant confounding. However, treatment estimates are biased if an individual’s outcome affects their sibling’s outcome. We propose a robustness test for assessing the presence of outcome-to-outcome interference in linear two-sibling FE models. We regress a gain-score—the difference between siblings' continuous outcomes—on both siblings' treatments and on a pre-treatment observed FE. Under certain restrictions, the observed FE’s partial regression coefficient signals the presence of outcome-to-outcome interference. Monte Carlo simulations demonstrated the robustness test under several models. We found that an observed FE signaled outcome-to-outcome spillover if it was directly associated with a sibling-invariant confounder of treatments and outcomes, directly associated with a sibling’s treatment, or directly and equally associated with both siblings’ outcomes. However, the robustness test collapsed if the observed FE was directly but differentially associated with siblings’ outcomes or if outcomes affected siblings’ treatments. 

\bigskip
\scriptsize{}

\noindent \textbf{Acknowledgements}: This work was supported by the Eunice Kennedy Shriver National Institute for Child Health and Human Development through the Population Research Center at the University of Texas-Austin (P2CHD042849) and through the Center for Demography at the University of Wisconsin-Madison (T32 HD007014-42). This content is solely the responsibility of the author and does not necessarily represent the official views of the National Institutes of Health. I thank Felix Elwert for comments and discussions that informed this manuscript. Conflicts of interest: none.

\small{}

\newpage

\tableofcontents

\newpage

\pagenumbering{arabic}
\setcounter{page}{1}


\section{Introduction}

Sibling fixed effects (FE) models are commonplace in observational epidemiologic research for estimating causal treatment effects while controlling for unobserved sibling-invariant confounding, such as shared genetic risk factors.\cite{frisell2012, khashan2014, khashan2015, hvolgaard2016, sjolander2016, hanley2017, axelsson2019, mallinson2020, petersen2020, frisell2021, mallinson2021} These models rely on strong assumptions—in particular, an individual's outcome cannot affect their sibling's treatment nor outcome.\cite{frisell2012, sjolander2016, frisell2021, mallinson2021} Such assumptions are precarious because siblings' health are interdependent.\cite{feinberg2012, deneve2017} One can prevent outcome-to-treatment interference by measuring all treatments before all outcomes within sibships. However, study design cannot circumvent outcome-to-outcome interference, thereby threatening the validity of any sibling FE analysis. Statistical tests like regressions with negative controls can assess potential bias,\cite{lipsitch2010, cunninghambook} although there is little guidance for their use in gauging outcome-to-outcome interference in sibling comparison studies.

We explain a robustness test for assessing outcome-to-outcome interference in linear two-sibling FE models. The method involves regressing a gain-score—the difference in siblings' outcomes—on both siblings' treatments and on a pre-treatment observed FE that is related to siblings’ treatments or outcomes. Using graphical causal modeling and Monte Carlo simulations, we demonstrate when the partial regression coefficient for the observed FE signals outcome-to-outcome interference across several models. 


\section{Methods}

\subsection{Baseline model} 

Our baseline sibling FE model (\textbf{Figure 1A}) reflects a typical sibling comparison design.\cite{sjolander2016} Subscripts $i=\left\{1,\ldots,N\right\}$ and $j=\left\{1,2\right\}$ denote cluster and sibling, respectively. $T_{ij}$ is a binary or continuous treatment, $Y_{ij}$ is a continuous outcome, $U_i$ is an unobserved FE (e.g., family health history), and $D_i=Y_{i2}-Y_{i1}$ is the gain-score. Arrows and adjacent Greek letters denote causal effects: $\delta$ is the treatment effect ($T_{ij}\rightarrow Y_{ij}$); $\psi$ is the unobserved confounding effect on outcomes ($U_i\rightarrow Y_{ij}$); and $\chi$ ($U_i\rightarrow T_{i1}$) and $\gamma$ ($U_i\rightarrow T_{i2}$) are unobserved confounding effects on treatments. This model embeds two assumptions. First, the treatment effect $\delta$ and the confounding effect $\psi$ are equal for both siblings. Second, all effects are linear and homogeneous. Both assumptions align with conventional linear FE models.\cite{sjolander2016, kim2019, kim2021, gunasekara2014,imai2019} For accessibility, we omitted sibling-specific covariates, which are compatible with the baseline model and all subsequent models as long as one can condition on them without loss of generality.

\afterpage{%
\begin{figure}
\centering
\resizebox{\linewidth}{!}{
\begin{tikzpicture}



\node (1) [scale=1.1] {$U_{i}$};

\node [above right =1.3cm and 2cm of 1, scale=1.1] (2) {$T_{i1}$};
\node [right =1.7cm of 2, scale=1.1] (3) {Y$_{i1}$};

\node [below right =1.3cm and 2cm of 1, scale=1.1] (4) {$T_{i2}$};
\node [right =1.7cm of 4, scale=1.1] (5) {$Y_{i2}$};

\node [right =7.4cm of 1, scale=1.1] (8) {$D_{i}$};

\node [above =2.8cm of 1, scale=1.3] (991) {\textbf{(1A)}};


\draw[Arrow] (1.east) -- (2.south west) node [midway, above, scale=1.1] {$\chi$};
\draw[Arrow3] (1) to [out=90, in=130] node [pos=0.6, above, scale=1.1] {$\psi$} (3);
\draw[Arrow3] (1) to [out=270, in=230] node [pos=0.6, below, scale=1.1] {$\psi$} (5);
\draw[Arrow] (1.east) -- (4.north west) node [midway, below, scale=1.1] {$\gamma$};

\draw[Arrow] (2.east) -- (3.west) node [midway, above, scale=1.1] {$\delta$};

\draw[Arrow] (4.east) -- (5.west) node [midway, above, scale=1.1] {$\delta$};

\draw[Arrow] (3.south east) -- (8.north west) node [midway, above, scale=1.1] {\it -1};
\draw[Arrow] (5.north east) -- (8.south west) node [midway, above, scale=1.1] {\it +1};




\node [right =10cm of 1, scale=1.1]  (21) {$U_{i}$};

\node [above=2.8cm of 21, scale=1.3] (993) {\textbf{(1B)}};

\node [above right =1.3cm and 2cm of 21, scale=1.1] (22) {$T_{i1}$};
\node [right =1.7cm of 22, scale=1.1] (23) {$Y_{i1}$};

\node [below right =1.3cm and 2cm of 21, scale=1.1] (24) {$T_{i2}$};
\node [right =1.7cm of 24, scale=1.1] (25) {$Y_{i2}$};

\node [right =7.4cm of 21, scale=1.1] (28) {$D_{i}$};


\draw[Arrow] (21.east) -- (22.south west) node [midway, above, scale=1.1] {$\chi$};
\draw[Arrow3] (21) to [out=90, in=130] node [pos=0.6, above, scale=1.1] {$\psi$} (23);
\draw[Arrow3] (21) to [out=270, in=230] node [pos=0.6, below, scale=1.1] {$\psi$} (25);
\draw[Arrow] (21.east) -- (24.north west) node [midway, below, scale=1.1] {$\gamma$};

\draw[Arrow] (22.east) -- (23.west) node [midway, above, scale=1.1] {$\delta$};

\draw[Arrow] (24.east) -- (25.west) node [midway, above, scale=1.1] {$\delta$};

\draw[Arrow] (23.south east) -- (28.north west) node [midway, above, scale=1.1] {\it -1};
\draw[Arrow] (25.north east) -- (28.south west) node [midway, above, scale=1.1] {\it +1};

\draw[Arrow] (23.south) -- (25.north) node [midway, right, scale=1.1] {$\eta$};




\node [below =9.5cm of 1, scale=1.1]  (41) {$U^\prime_{i}$};

\node [above =2.8cm of 41, scale=1.3] (995) {\textbf{(1C)}};

\node [above right =1.3cm and 2cm of 41, scale=1.1] (42) {$T_{i1}$};
\node [right =1.7cm of 42, scale=1.1] (43) {$Y_{i1}$};

\node [below right =1.3cm and 2cm of 41, scale=1.1] (44) {$T_{i2}$};
\node [right =1.7cm of 44, scale=1.1] (45) {$Y_{i2}$};

\node [right =1.5cm of 41, scale=1.1] (46) {$C_{i}$};

\node [right =7.4cm of 41, scale=1.1] (48) {$D_{i}$};


\draw[Arrow] (41.east) -- (42.south west) node [midway, above, scale=1.1] {$\chi$};
\draw[Arrow3] (41) to [out=90, in=130] node [pos=0.6, above, scale=1.1] {$\psi$} (43);
\draw[Arrow3] (41) to [out=270, in=230] node [pos=0.6, below, scale=1.1] {$\psi$} (45);
\draw[Arrow] (41.east) -- (44.north west) node [midway, below, scale=1.1] {$\gamma$};

\draw[Arrow] (42.east) -- (43.west) node [midway, above, scale=1.1] {$\delta$};

\draw[Arrow] (44.east) -- (45.west) node [midway, above, scale=1.1] {$\delta$};

\draw[Arrow] (43.south east) -- (48.north west) node [midway, above, scale=1.1] {\it -1};
\draw[Arrow] (45.north east) -- (48.south west) node [midway, above, scale=1.1] {\it +1};

\draw[dashed, thick] (41.east)--(46.west) node [midway, above, scale=1.1] {$\pi$};




\node [below =9.5cm of 21, scale=1.1]  (61) {$U^\prime_{i}$};

\node [above =2.8cm of 61, scale=1.3] (997) {\textbf{(1D)}};

\node [above right =1.3cm and 2cm of 61, scale=1.1] (62) {$T_{i1}$};
\node [right =1.7cm of 62, scale=1.1] (63) {$Y_{i1}$};

\node [below right =1.3cm and 2cm of 61, scale=1.1] (64) {$T_{i2}$};
\node [right =1.7cm of 64, scale=1.1] (65) {$Y_{i2}$};

\node [right =1.5cm of 61, scale=1.1] (66) {$C_{i}$};

\node [right =7.4cm of 61, scale=1.1] (68) {$D_{i}$};


\draw[Arrow] (61.east) -- (62.south west) node [midway, above, scale=1.1] {$\chi$};
\draw[Arrow3] (61) to [out=90, in=130] node [pos=0.6, above, scale=1.1] {$\psi$} (63);
\draw[Arrow3] (61) to [out=270, in=230] node [pos=0.6, below, scale=1.1] {$\psi$} (65);
\draw[Arrow] (61.east) -- (64.north west) node [midway, below, scale=1.1] {$\gamma$};

\draw[Arrow] (62.east) -- (63.west) node [midway, above, scale=1.1] {$\delta$};

\draw[Arrow] (64.east) -- (65.west) node [midway, above, scale=1.1] {$\delta$};

\draw[Arrow] (63.south east) -- (68.north west) node [midway, above, scale=1.1] {\it -1};
\draw[Arrow] (65.north east) -- (68.south west) node [midway, above, scale=1.1] {\it +1};

\draw[dashed, thick] (61.east)--(66.west) node [midway, above, scale=1.1] {$\pi$};

\draw[Arrow] (63.south) -- (65.north) node [midway, right, scale=1.1] {$\eta$};

\end{tikzpicture}
}
\caption{\textbf{Figure 1.} Graphical causal models of sibling fixed effects models. Subscripts $i$ and $j$ denote cluster and sibling, respectively. $T_{ij}$ is the exposure, $Y_{ij}$ is the outcome, $D_i$ is the gain-score, $U_{i}$ and $U^\prime_{i}$ are unobserved fixed effects, and $C_{i}$ is a pre-treatment observed fixed effect. Arrows indicate direct and linear causal effects, and dashed lines indicate linear associations. Greek letters denote the magnitude of an effect or association. (1A) and (1C) do not have outcome-to-outcome interference, whereas (1B) and (1D) have outcome-to-outcome interference.} 
\end{figure}
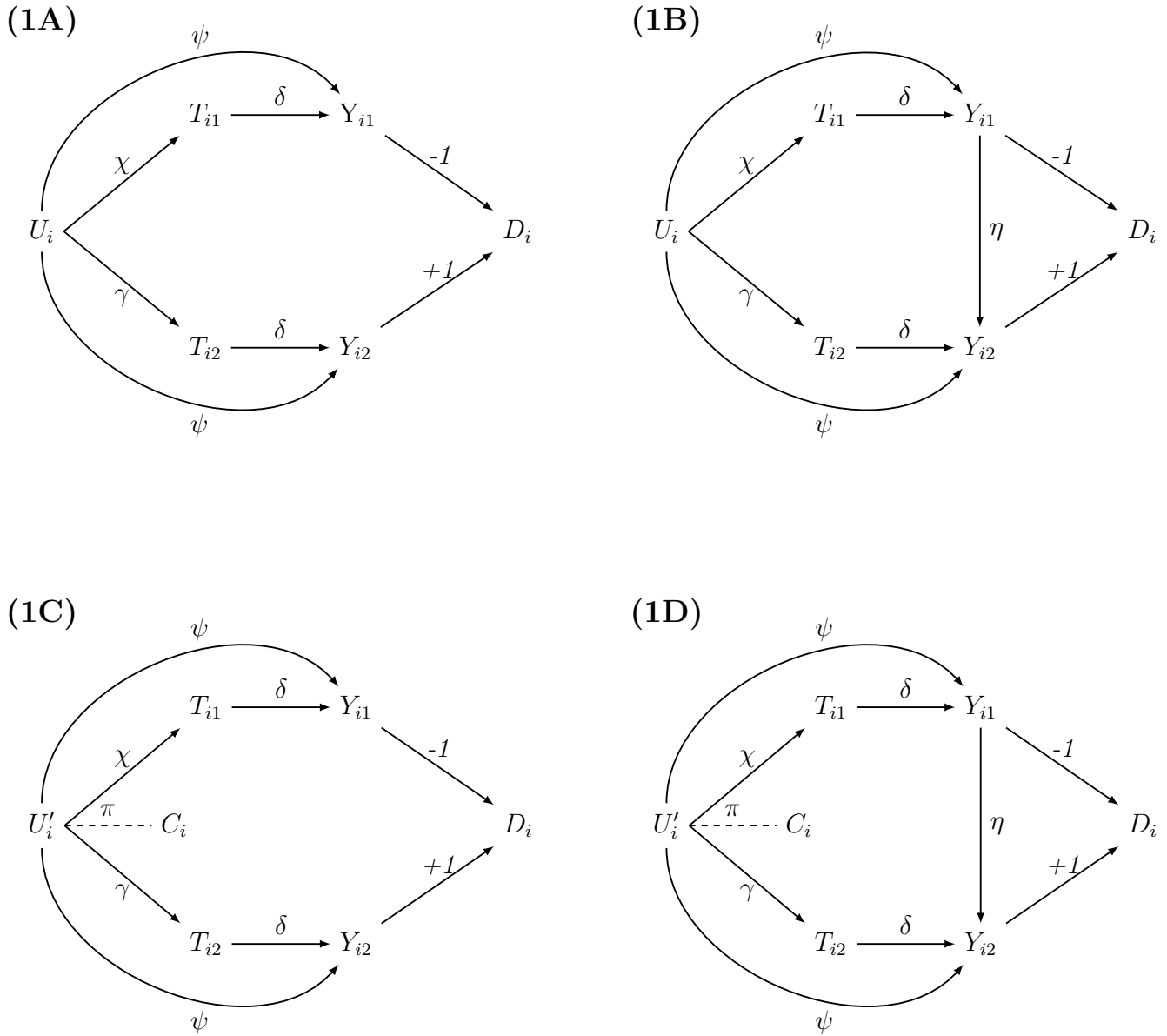
\clearpage
}

\subsection{Gain-score regression} 

Following the baseline model, $\delta$ is our estimand. Regressing $Y_{ij}$ on $T_{ij}$ will not identify $\delta$, covariate adjustment notwithstanding, because $U_i$ confounds both variables. This motivates the utility of FE estimators that isolate treatment effects while controlling for certain types of unobserved confounding.\cite{sjolander2016, mallinson2021,kim2019,kim2021,gunasekara2014,imai2019} The gain-score is one such estimator, and we demonstrate its mechanics and susceptibility to outcome-to-outcome interference in simple linear regression. We regress the gain-score on siblings’ treatments, 
\begin{equation}
D_i=b_1T_{i1}+b_2T_{i2}+e_i,
\end{equation}

\noindent where $e_i$ is the residual. Through outcome differencing, the gain-score estimator offsets unobserved confounding bias such that the treatment-conditional covariance between $U_i$ and $D_i$ ($\sigma_{U_iD_i\cdot T_{i1}T_{i2}}$) is zero, thereby isolating the treatment effect.\cite{mallinson2021, kim2019, kim2021} The baseline model includes four pathways that “transmit” bias from $U_i$ into $D_i$: 
\begin{enumerate}
\item $U_{i}\rightarrow T_{i1}\rightarrow Y_{i1}\rightarrow D_i$: $-\chi\delta$
\item $U_{i}\rightarrow T_{i2}\rightarrow Y_{i2}\rightarrow D_{i}$: $\gamma\delta$
\item $U_{i}\rightarrow Y_{i1}\rightarrow D_{i}$: $-\psi$
\item $U_{i}\rightarrow Y_{i2}\rightarrow D_{i}$: $\psi$
\end{enumerate}

\noindent Covariate adjustment on $T_{i1}$ and $T_{i2}$ closes paths 1 and 2, and paths 3 and 4 cancel each other out exactly because the association that flows along path 3 ($-\psi$) is equal to the negative value of the association that flows along path 4 ($\psi$). Thus, $\sigma_{U_iD_i\cdot T_{i1}T_{i2}}=0$, and $b_2=\delta$ identifies the treatment effect precisely (likewise, $b_1=-\delta$). See \textbf{Appendix} for details.

The estimator’s validity relies on absent outcome-to-outcome interference. Otherwise, we cannot identify $\delta$. The model in \textbf{Figure 1B} introduces outcome-to-outcome interference $\eta$ ($Y_{i1}\rightarrow Y_{i2}$). There are now two additional pathways from $U_i$ to $D_i$:
\begin{enumerate}
\setcounter{enumi}{4}
\item $U_{i}\rightarrow T_{i1}\rightarrow Y_{i1}\rightarrow Y_{i2}\rightarrow D_i$: $\chi\delta\eta$
\item $U_{i}\rightarrow Y_{i1}\rightarrow Y_{i2}\rightarrow D_{i}$: $\psi\eta$
\end{enumerate}

Covariate adjustment on $T_{i1}$ blocks path 5. However, path 6 is open as a result of outcome-to-outcome interference, so $\sigma_{U_iD_i\cdot T_{i1}T_{i2}}\neq0$, thereby preventing treatment identification (i.e., $b_2\neq\delta$ and $b_1\neq-\delta$). See \textbf{Appendix} for details.

\subsection{Robustness test}

Given that outcome-to-outcome interference in sibling FE models biases treatment identification, we may consider statistical tests for gauging this source of bias. An ideal test is a gain-score regression that adjusts for $U_i$:
\begin{equation}
D_i=b_1T_{i1}+b_2T_{i2}+b_{U}U_{i}+e_i.
\end{equation}

\noindent Under the models in \textbf{Figures 1A and 1B}, $b_U=0$ indicates no outcome-to-outcome spillover ($\eta=0$) because $\sigma_{U_iD_i\cdot T_{i1}T_{i2}}=0$, whereas $b_U\neq0$ indicates outcome-to-outcome spillover ($\eta\neq0$) because $\sigma_{U_iD_i\cdot T_{i1}T_{i2}}\neq0$. This hypothetical test is straightforward but impossible; $U_i$ is unobserved, so we cannot adjust for it. Indeed, adjustment for $U_i$ would render sibling FE analysis obsolete—simply regress $Y_{ij}$ on $T_{ij}$ and $U_i$ to identify $\delta$.  

Still, we may measure cluster-level characteristics, raising the possibility of adjusting for an observed FE in the gain-score regression. The model in \textbf{Figure 1C} introduces $C_i$, a binary or continuous observed pre-treatment FE, and $U_i^\prime$, an unobserved FE that is equivalent to $U_i$ without $C_i$. The dashed line between $U_i^\prime$ and $C_i$ indicates a direct and linear association that is equal to $\pi$. An example of a potential observed FE is maternal nativity, which is a sibling-invariant characteristic that is associated with family health history and pregnancy-related outcomes in the United States.\cite{qin2010, creanga2012} Additionally, $C_i$ should precede siblings’ treatments to prevent bias from conditioning on a mediator variable between the treatment and outcome or from conditioning on a post-treatment collider.\cite{elwert2014, vwbook}

Under the model in \textbf{Figure 1C}, we repeat the gain-score regression and adjust for $C_i$: 
\begin{equation}
D_i=b_1T_{i1}+b_2T_{i2}+b_{C}C_{i}+e_i.
\end{equation}

\noindent There are four pathways from $C_i$ to $D_{i}$: 
\begin{enumerate}
\item $C_{i}\hdashrule[0.5ex]{0.6cm}{0.5pt}{2pt} U^\prime_{i}\rightarrow T_{i1}\rightarrow Y_{i1}\rightarrow D_i$: $-\pi\chi\delta$
\item $C_{i}\hdashrule[0.5ex]{0.6cm}{0.5pt}{2pt} U^\prime_{i}\rightarrow T_{i2}\rightarrow Y_{i2}\rightarrow D_{i}$: $\pi\gamma\delta$
\item $C_{i}\hdashrule[0.5ex]{0.6cm}{0.5pt}{2pt} U^\prime_{i}\rightarrow Y_{i1}\rightarrow D_{i}$: $-\pi\psi$
\item $C_{i}\hdashrule[0.5ex]{0.6cm}{0.5pt}{2pt} U^\prime_{i}\rightarrow Y_{i2}\rightarrow D_{i}$: $\pi\psi$
\end{enumerate}

\noindent Covariate adjustment on siblings’ treatments blocks paths 1 and 2, and outcome differencing offsets the associations from paths 3 and 4. Therefore, $\sigma_{C_iD_i\cdot T_{i1}T_{i2}}=0$, so $b_C=0$.

We then consider the model in \textbf{Figure 1D}, which includes $C_i$ and outcome-to-outcome interference. Two additional pathways link $C_i$ to $D_i$: 

\begin{enumerate}
\setcounter{enumi}{4}
\item $C_{i}\hdashrule[0.5ex]{0.6cm}{0.5pt}{2pt} U^\prime_{i}\rightarrow T_{i1}\rightarrow Y_{i1}\rightarrow Y_{i2}\rightarrow D_i$: $\chi\delta\eta$
\item $C_{i}\hdashrule[0.5ex]{0.6cm}{0.5pt}{2pt} U^\prime_{i}\rightarrow Y_{i1}\rightarrow Y_{i2}\rightarrow D_{i}$: $\pi\psi\eta$
\end{enumerate}

\noindent While covariate adjustment closes path 5, path 6 remains open, so $\sigma_{C_iD_i\cdot T_{i1}T_{i2}}\neq0$ and $b_C\neq0$. Theoretically, this robustness test signals outcome-to-outcome interference in a basic sibling FE model. We now turn to simulation analyses to exercise this test in multiple settings. 


\section{Simulations}

Monte Carlo simulations demonstrated the robustness test’s performance in the model of \textbf{Figure 1D} and across its subsequent variations in \textbf{Figure 2}.\cite{adkins2012} Model alterations included treatment-to-treatment interference, direct associations between $C_i$ and siblings’ treatments or outcomes, treatment-to-outcome interference, or outcome-to-treatment interference.

\afterpage{%
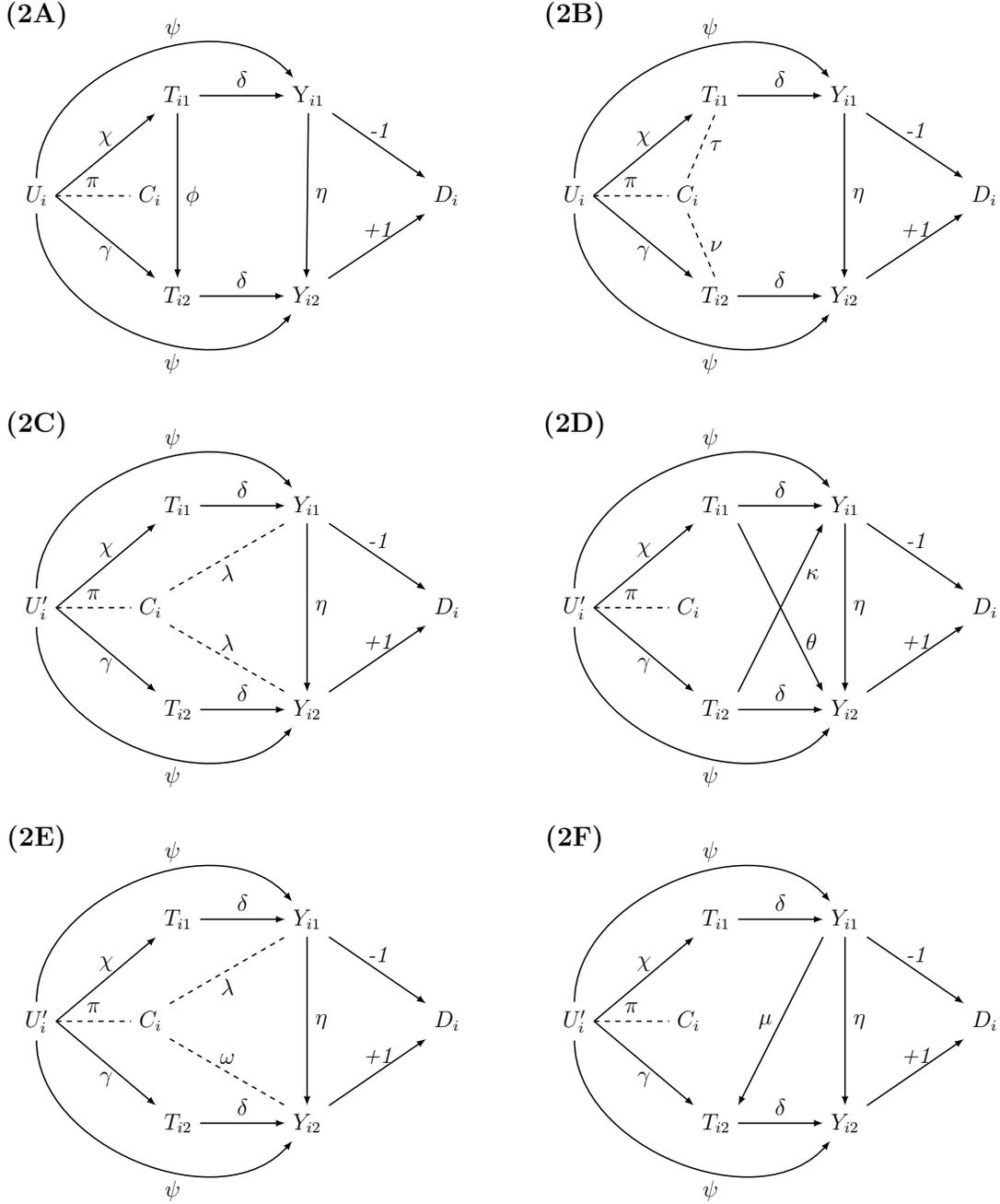
\begin{figure}
\centering
\resizebox{!}{7in}{
\begin{tikzpicture}



\node (1) [scale=1.1] {$U_{i}$};

\node [above right =1.3cm and 2cm of 1, scale=1.1] (2) {$T_{i1}$};
\node [right =1.7cm of 2, scale=1.1] (3) {Y$_{i1}$};

\node [below right =1.3cm and 2cm of 1, scale=1.1] (4) {$T_{i2}$};
\node [right =1.7cm of 4, scale=1.1] (5) {$Y_{i2}$};

\node [right =7.4cm of 1, scale=1.1] (8) {$D_{i}$};

\node [right =1.5cm of 1, scale=1.1] (6) {$C_{i}$};

\node [above =2.8cm of 1, scale=1.3] (991) {\textbf{(2A)}};


\draw[Arrow] (1.east) -- (2.south west) node [midway, above, scale=1.1] {$\chi$};
\draw[Arrow3] (1) to [out=90, in=130] node [pos=0.6, above, scale=1.1] {$\psi$} (3);
\draw[Arrow3] (1) to [out=270, in=230] node [pos=0.6, below, scale=1.1] {$\psi$} (5);
\draw[Arrow] (1.east) -- (4.north west) node [midway, below, scale=1.1] {$\gamma$};

\draw[Arrow] (2.east) -- (3.west) node [midway, above, scale=1.1] {$\delta$};

\draw[Arrow] (4.east) -- (5.west) node [midway, above, scale=1.1] {$\delta$};

\draw[Arrow] (3.south east) -- (8.north west) node [midway, above, scale=1.1] {\it -1};
\draw[Arrow] (5.north east) -- (8.south west) node [midway, above, scale=1.1] {\it +1};

\draw[dashed, thick] (1.east)--(6.west) node [midway, above, scale=1.1] {$\pi$};

\draw[Arrow] (3.south) -- (5.north) node [midway, right, scale=1.1] {$\eta$};

\draw[Arrow] (2.south)--(4.north) node [midway, right, scale=1.1] {$\phi$};




\node [right =10cm of 1, scale=1.1]  (21) {$U_{i}$};

\node [above=2.8cm of 21, scale=1.3] (993) {\textbf{(2B)}};

\node [above right =1.3cm and 2cm of 21, scale=1.1] (22) {$T_{i1}$};
\node [right =1.7cm of 22, scale=1.1] (23) {$Y_{i1}$};

\node [below right =1.3cm and 2cm of 21, scale=1.1] (24) {$T_{i2}$};
\node [right =1.7cm of 24, scale=1.1] (25) {$Y_{i2}$};

\node [right =7.4cm of 21, scale=1.1] (28) {$D_{i}$};

\node [right =1.5cm of 21, scale=1.1] (26) {$C_{i}$};


\draw[Arrow] (21.east) -- (22.south west) node [midway, above, scale=1.1] {$\chi$};
\draw[Arrow3] (21) to [out=90, in=130] node [pos=0.6, above, scale=1.1] {$\psi$} (23);
\draw[Arrow3] (21) to [out=270, in=230] node [pos=0.6, below, scale=1.1] {$\psi$} (25);
\draw[Arrow] (21.east) -- (24.north west) node [midway, below, scale=1.1] {$\gamma$};

\draw[Arrow] (22.east) -- (23.west) node [midway, above, scale=1.1] {$\delta$};

\draw[Arrow] (24.east) -- (25.west) node [midway, above, scale=1.1] {$\delta$};

\draw[Arrow] (23.south east) -- (28.north west) node [midway, above, scale=1.1] {\it -1};
\draw[Arrow] (25.north east) -- (28.south west) node [midway, above, scale=1.1] {\it +1};

\draw[dashed, thick] (21.east)--(26.west) node [midway, above, scale=1.1] {$\pi$};

\draw[Arrow] (23.south) -- (25.north) node [midway, right, scale=1.1] {$\eta$};

\draw[dashed, thick]  (26.north)--(22.south) node [midway, right, scale=1.1] {$\tau$};
\draw[dashed, thick]  (26.south)--(24.north) node [midway, right, scale=1.1] {$\nu$};




\node [below =7.5cm of 1, scale=1.1]  (41) {$U^\prime_{i}$};

\node [above =2.8cm of 41, scale=1.3] (995) {\textbf{(2C)}};

\node [above right =1.3cm and 2cm of 41, scale=1.1] (42) {$T_{i1}$};
\node [right =1.7cm of 42, scale=1.1] (43) {$Y_{i1}$};

\node [below right =1.3cm and 2cm of 41, scale=1.1] (44) {$T_{i2}$};
\node [right =1.7cm of 44, scale=1.1] (45) {$Y_{i2}$};

\node [right =1.5cm of 41, scale=1.1] (46) {$C_{i}$};

\node [right =7.4cm of 41, scale=1.1] (48) {$D_{i}$};


\draw[Arrow] (41.east) -- (42.south west) node [midway, above, scale=1.1] {$\chi$};
\draw[Arrow3] (41) to [out=90, in=130] node [pos=0.6, above, scale=1.1] {$\psi$} (43);
\draw[Arrow3] (41) to [out=270, in=230] node [pos=0.6, below, scale=1.1] {$\psi$} (45);
\draw[Arrow] (41.east) -- (44.north west) node [midway, below, scale=1.1] {$\gamma$};

\draw[Arrow] (42.east) -- (43.west) node [midway, above, scale=1.1] {$\delta$};

\draw[Arrow] (44.east) -- (45.west) node [midway, above, scale=1.1] {$\delta$};

\draw[Arrow] (43.south east) -- (48.north west) node [midway, above, scale=1.1] {\it -1};
\draw[Arrow] (45.north east) -- (48.south west) node [midway, above, scale=1.1] {\it +1};

\draw[dashed, thick] (41.east)--(46.west) node [midway, above, scale=1.1] {$\pi$};

\draw[Arrow] (43.south) -- (45.north) node [midway, right, scale=1.1] {$\eta$};

\draw[dashed, thick]  (46.north east)--(43.south west) node [midway, below, scale=1.1] {$\lambda$};
\draw[dashed, thick]  (46.south east)--(45.north west) node [midway, above, scale=1.1] {$\lambda$};



\node [below =7.5cm of 21, scale=1.1]  (71) {$U^\prime_{i}$};

\node [above =2.8cm of 71, scale=1.3] (998) {\textbf{(2D)}};

\node [above right =1.3cm and 2cm of 71, scale=1.1] (72) {$T_{i1}$};
\node [right =1.7cm of 72, scale=1.1] (73) {$Y_{i1}$};

\node [below right =1.3cm and 2cm of 71, scale=1.1] (74) {$T_{i2}$};
\node [right =1.7cm of 74, scale=1.1] (75) {$Y_{i2}$};

\node [right =1.5cm of 71, scale=1.1] (76) {$C_{i}$};

\node [right =7.4cm of 71, scale=1.1] (78) {$D_{i}$};


\draw[Arrow] (71.east) -- (72.south west) node [midway, above, scale=1.1] {$\chi$};
\draw[Arrow3] (71) to [out=90, in=130] node [pos=0.6, above, scale=1.1] {$\psi$} (73);
\draw[Arrow3] (71) to [out=270, in=230] node [pos=0.6, below, scale=1.1] {$\psi$} (75);
\draw[Arrow] (71.east) -- (74.north west) node [midway, below, scale=1.1] {$\gamma$};

\draw[Arrow] (72.east) -- (73.west) node [midway, above, scale=1.1] {$\delta$};

\draw[Arrow] (74.east) -- (75.west) node [midway, above, scale=1.1] {$\delta$};

\draw[Arrow] (73.south east) -- (78.north west) node [midway, above, scale=1.1] {\it -1};
\draw[Arrow] (75.north east) -- (78.south west) node [midway, above, scale=1.1] {\it +1};

\draw[dashed, thick] (71.east)--(76.west) node [midway, above, scale=1.1] {$\pi$};

\draw[Arrow] (73.south) -- (75.north) node [midway, right, scale=1.1] {$\eta$};

\draw[Arrow] (72.south east)--(75.north west) node [pos=0.7, right, scale=1.1] {$\theta$};
\draw[Arrow] (74.north east)--(73.south west) node [pos=0.7, right, scale=1.1] {$\kappa$};



\node [below =7.5cm of 41, scale=1.1]  (61) {$U^\prime_{i}$};

\node [above =2.8cm of 61, scale=1.3] (997) {\textbf{(2E)}};

\node [above right =1.3cm and 2cm of 61, scale=1.1] (62) {$T_{i1}$};
\node [right =1.7cm of 62, scale=1.1] (63) {$Y_{i1}$};

\node [below right =1.3cm and 2cm of 61, scale=1.1] (64) {$T_{i2}$};
\node [right =1.7cm of 64, scale=1.1] (65) {$Y_{i2}$};

\node [right =1.5cm of 61, scale=1.1] (66) {$C_{i}$};

\node [right =7.4cm of 61, scale=1.1] (68) {$D_{i}$};


\draw[Arrow] (61.east) -- (62.south west) node [midway, above, scale=1.1] {$\chi$};
\draw[Arrow3] (61) to [out=90, in=130] node [pos=0.6, above, scale=1.1] {$\psi$} (63);
\draw[Arrow3] (61) to [out=270, in=230] node [pos=0.6, below, scale=1.1] {$\psi$} (65);
\draw[Arrow] (61.east) -- (64.north west) node [midway, below, scale=1.1] {$\gamma$};

\draw[Arrow] (62.east) -- (63.west) node [midway, above, scale=1.1] {$\delta$};

\draw[Arrow] (64.east) -- (65.west) node [midway, above, scale=1.1] {$\delta$};

\draw[Arrow] (63.south east) -- (68.north west) node [midway, above, scale=1.1] {\it -1};
\draw[Arrow] (65.north east) -- (68.south west) node [midway, above, scale=1.1] {\it +1};

\draw[dashed, thick] (61.east)--(66.west) node [midway, above, scale=1.1] {$\pi$};

\draw[Arrow] (63.south) -- (65.north) node [midway, right, scale=1.1] {$\eta$};

\draw[dashed, thick]  (66.north east)--(63.south west) node [midway, below, scale=1.1] {$\lambda$};
\draw[dashed, thick]  (66.south east)--(65.north west) node [midway, above, scale=1.1] {$\omega$};



\node [below =7.5cm of 71, scale=1.1]  (81) {$U^\prime_{i}$};

\node [above =2.8cm of 81, scale=1.3] (999) {\textbf{(2F)}};

\node [above right =1.3cm and 2cm of 81, scale=1.1] (82) {$T_{i1}$};
\node [right =1.7cm of 82, scale=1.1] (83) {$Y_{i1}$};

\node [below right =1.3cm and 2cm of 81, scale=1.1] (84) {$T_{i2}$};
\node [right =1.7cm of 84, scale=1.1] (85) {$Y_{i2}$};

\node [right =1.5cm of 81, scale=1.1] (86) {$C_{i}$};

\node [right =7.4cm of 81, scale=1.1] (88) {$D_{i}$};


\draw[Arrow] (81.east) -- (82.south west) node [midway, above, scale=1.1] {$\chi$};
\draw[Arrow3] (81) to [out=90, in=130] node [pos=0.6, above, scale=1.1] {$\psi$} (83);
\draw[Arrow3] (81) to [out=270, in=230] node [pos=0.6, below, scale=1.1] {$\psi$} (85);
\draw[Arrow] (81.east) -- (84.north west) node [midway, below, scale=1.1] {$\gamma$};

\draw[Arrow] (82.east) -- (83.west) node [midway, above, scale=1.1] {$\delta$};

\draw[Arrow] (84.east) -- (85.west) node [midway, above, scale=1.1] {$\delta$};

\draw[Arrow] (83.south east) -- (88.north west) node [midway, above, scale=1.1] {\it -1};
\draw[Arrow] (85.north east) -- (88.south west) node [midway, above, scale=1.1] {\it +1};

\draw[dashed, thick] (81.east)--(86.west) node [midway, above, scale=1.1] {$\pi$};

\draw[Arrow] (83.south) -- (85.north) node [midway, right, scale=1.1] {$\eta$};

\draw[Arrow] (83.south west)--(84.north east) node [midway, left, scale=1.1] {$\mu$};

\end{tikzpicture}
}
\caption{\textbf{Figure 2.} Graphical causal models of various sibling fixed effects models with an observed fixed effect. Subscripts $i$ and $j$ denote cluster and sibling, respectively. $T_{ij}$ is the exposure, $Y_{ij}$ is the outcome, $D_i$ is the gain-score, $U_{i}$ and $U^\prime_{i}$ are unobserved fixed effects, and $C_{i}$ is a pre-treatment observed fixed effect. Arrows indicate direct and linear causal effects, and dashed lines indicate linear associations. Greek letters denote the magnitude of an effect or association.}
\end{figure}
\clearpage
}

Our simulation model follows:

\[U^\prime_{i},\upsilon_{c},\upsilon_{i1},\upsilon_{i2}\sim N(0,1)\]
\[\boldsymbol\alpha=\{\lambda,\omega\}\]
\[C_{i}=
	\begin{cases}
	0 \ \text{if}\ \pi U^\prime_{i} + \upsilon_{c}\leq 1 \\
	1 \ \text{if}\ \pi U^\prime_{i} + \upsilon_{c} > 1
	\end{cases}\]
\[T_{i1}=
	\begin{cases}
	0 \ \text{if}\ \tau C_{i}+\chi U^\prime_{i}\leq -0.2 \\
	1 \ \text{if}\ \tau C_{i}+\chi U^\prime_{i} > -0.2
	\end{cases}\]
\[T_{i2}=
	\begin{cases}
	0 \ \text{if}\ \phi T_{i1}+\mu Y_{i1}+\nu C_{i}+\gamma U^\prime_{i}\leq 1 \\
	1 \ \text{if}\ \phi T_{i1}+\mu Y_{i1}+\nu C_{i}+\gamma U^\prime_{i} > 1
	\end{cases}\]
\[Y_{i1}=\delta T_{i1}+\kappa T_{i2}+\lambda C_{i}+\psi U^\prime_{i}+\upsilon_{i1}\]
\[Y_{i2}=\delta T_{i2}+\theta T_{i1}+\eta Y_{i1}+\boldsymbol\alpha C_{i}+\psi U^\prime_{i}+\upsilon_{i2}\]
\[D_i=Y_{i2}-Y_{i1}\]

For each sibling FE model, we ran two simulation sets, each consisting of two simulations with 1,000 runs of 5,000 observations (sibling pairs). Simulations varied on outcome-to-outcome interference ($\eta$) and on the association between $C_i$ and $U_i^\prime$ ($\pi$): 

\begin{enumerate} 
\item Simulation 1.1: $\eta=0$ and $\pi=0.5$; 
\item Simulation 1.2: $\eta=0.3$ and $\pi=0.5$; 
\item Simulation 2.1: $\eta=0$ and $\pi=0$; 
\item Simulation 2.2: $\eta=0.3$ and $\pi=0$. 
\end{enumerate}

Baseline parameters were fixed across simulations: $\delta=3$, $\chi=1$, $\gamma=2$, and $\psi=5$. Unless specified, we set additional parameters to zero. Within samples, we ran the gain-score regression in equation (3). We conducted simulations in Stata Statistical Software: Release 16.\cite{statasoft} See \textbf{Appendix} for simulation code.

\textbf{Table 1} displays simulation inputs and results for the observed FE, and \textbf{Appendix Table 1} contains full results. We first consider simulations 1.1 and 1.2. Under the model of \textbf{Figure 1D}, $\widehat{b_C}=0$ (95\% confidence interval: -0.105, 0.105) without outcome-to-outcome interference, and the coverage probability—the percent of samples with confidence intervals that overlapped zero—exceeded 95\%. Conversely, $\widehat{b_C}=0.264$ (95\% confidence interval: 0.159, 0.370) with outcome-to-outcome interference, and the coverage probability was \textless1\%. These results persisted in the presence of treatment-to-treatment interference (\textbf{Figure 2A}), direct associations between $C_i$ and siblings’ treatments (\textbf{Figure 2B}), equivalent  and direct associations between $C_i$ and siblings’ outcomes (\textbf{Figure 2C}), and treatment-to-outcome interference (\textbf{Figure 2D}). However, $\widehat{b_C}\neq0$ with direct non-equivalent associations between $C_i$ and siblings’ outcomes (\textbf{Figure 2E}) or with outcome-to-treatment interference (\textbf{Figure 2F}), regardless if outcome-to-outcome interference was present. 

\afterpage{%
\begin{landscape}
\begin{onehalfspacing}
\begin{footnotesize}
\centering
\setlength\LTleft{0pt}
\setlength\LTright{0pt}
\begin{longtable}{l l S[table-format=1.3] c S[table-format=2.1] S[table-format=1.3] c S[table-format=2.1] c}
\captionsetup{justification=justified, singlelinecheck=false}
\caption{\textbf{Table 1.} Monte Carlo simulation results} \\
\toprule[1pt]\midrule[0.3pt] 
{} & {} & \multicolumn{3}{c}{\textbf{Simulation Set 1.1}} & \multicolumn{3}{c}{\textbf{Simulation Set 1.2}} & {} \\
\textbf{Model} & \textbf{Additional Parameters} & \multicolumn{3}{c}{\textbf{$\eta=0$, $\pi=0.5$}} & \multicolumn{3}{c}{\textbf{$\eta=0.3$, $\pi=0.5$}} & \textbf{SS1:} \\ \cline{3-5} \cline{6-8}
\textbf{(Fig.)} & \textbf{(With Baseline)} & \textbf{\textit{$b_C$}} & \textbf{\textit{95\% CI}} & \textbf{\textit{Cover (\%)}}$^a$ & \textbf{\textit{$b_C$}} & \textbf{\textit{95\% CI}} & \textbf{\textit{Cover (\%)}}$^a$ & \textbf{Test Valid}$^{b}$ \\
\hline 
\rowcolor{lightgray}
\textbf{1D} & None & -0.000 & -0.105, 0.105 & 95.1 & 0.264 & 0.159, 0.370 & 0.3 & Yes \\
\textbf{2A} & $\phi=0.9$ ($T_{i1}\rightarrow T_{i2}$) & -0.000 & -0.105, 0.104 & 95.5 & 0.397 & 0.289, 0.505 & 0.0 & Yes \\
\rowcolor{lightgray}
\textbf{2B} & $\tau=0.7$ ($C_{i}\hdashrule[0.5ex]{0.6cm}{0.5pt}{2pt} T_{i1}$), $\nu=1.5$ ($C_{i}\hdashrule[0.5ex]{0.6cm}{0.5pt}{2pt} T_{i2}$) & -0.001 & -0.116, 0.114 & 94.5 & -0.378 & -0.496, -0.260 & 0.0 & Yes \\
\textbf{2C} & $\lambda=2.5$ ($C_{i}\hdashrule[0.5ex]{0.6cm}{0.5pt}{2pt} Y_{ij}$) & -0.000 & -0.105, 0.105 & 95.1 & 1.014 & 0.909, 1.120 & 0.0 & Yes \\
\rowcolor{lightgray}
\textbf{2D} & $\theta=1.2$ ($T_{i1}\rightarrow Y_{i2}$), $\kappa=0.6$ ($T_{i2}\rightarrow Y_{i1}$) & -0.000 & -0.105, 0.105 & 95.1 & 1.314 & 1.209, 1.420 & 0.0 & Yes \\
\textbf{2E} & $\lambda=2.5$ ($C_{i}\hdashrule[0.5ex]{0.6cm}{0.5pt}{2pt} Y_{i1}$), $\omega=2.8$ ($C_{i}\hdashrule[0.5ex]{0.6cm}{0.5pt}{2pt} Y_{i2}$) &  0.300 & 0.195, 0.405 & 0.0 & 0.264 & 0.159, 0.370 & 0.3 & No \\
\rowcolor{lightgray}
\textbf{2F} & $\mu=0.4$ ($Y_{i1}\rightarrow T_{i2}$) & 0.033 & -0.071, 0.136 & 89.3 & 0.462 & 0.351, 0.573 & 0.0 & No \\
\hline 
{} & {} & \multicolumn{3}{c}{\textbf{Simulation Set 2.1}} & \multicolumn{3}{c}{\textbf{Simulation Set 2.2}} & {} \\
\textbf{Model} & \textbf{Additional Parameters} & \multicolumn{3}{c}{\textbf{$\eta=0$, $\pi=0$}} & \multicolumn{3}{c}{\textbf{$\eta=0.3$, $\pi=0$}} & \textbf{SS2:} \\ \cline{3-5} \cline{6-8}
\textbf{(Fig.)} & \textbf{(With Baseline)} & \textbf{\textit{$b_C$}} & \textbf{\textit{95\% CI}} & \textbf{\textit{Cover (\%)}}$^a$ & \textbf{\textit{$b_C$}} & \textbf{\textit{95\% CI}} & \textbf{\textit{Cover (\%)}}$^a$ & \textbf{Test Valid}$^{b}$ \\\rowcolor{lightgray}
\textbf{1D} & None & -0.002 & -0.109, 0.105 & 94.0 & -0.002 & -0.109, 0.106 & 93.1 & No \\
\textbf{2A} & $\phi=0.9$ ($T_{i1}\rightarrow T_{i2}$) & -0.002 & -0.109, 0.105 & 94.1 & -0.001 & -0.113, 0.111 & 93.1 & No \\
\rowcolor{lightgray}
\textbf{2B} & $\tau=0.7$ ($C_{i}\hdashrule[0.5ex]{0.6cm}{0.5pt}{2pt} T_{i1}$), $\nu=1.5$ ($C_{i}\hdashrule[0.5ex]{0.6cm}{0.5pt}{2pt} T_{i2}$) & -0.002 & -0.112, 0.108 & 94.2 & -0.825 & -0.936, -0.714 & 0.0 & Yes \\
\textbf{2C} & $\lambda=2.5$ ($C_{i}\hdashrule[0.5ex]{0.6cm}{0.5pt}{2pt} Y_{ij}$) & -0.002 & -0.109, 0.105 & 94.0 & 0.749 & 0.641, 0.856 & 0.0 & Yes \\
\rowcolor{lightgray}
\textbf{2D} & $\theta=1.2$ ($T_{i1}\rightarrow Y_{i2}$), $\kappa=0.6$ ($T_{i2}\rightarrow Y_{i1}$) & -0.002 & -0.109, 0.105 & 94.0 & 1.049 & 0.941, 1.156 & 0.0 & No \\
\textbf{2E} & $\lambda=2.5$ ($C_{i}\hdashrule[0.5ex]{0.6cm}{0.5pt}{2pt} Y_{i1}$), $\omega=2.8$ ($C_{i}\hdashrule[0.5ex]{0.6cm}{0.5pt}{2pt} Y_{i2}$) & 0.298 & 0.191, 0.405 & 0.1 & 1.049 & 0.941, 1.156 & 0.0 & No \\
\rowcolor{lightgray}
\textbf{2F} & $\mu=0.4$ ($Y_{i1}\rightarrow T_{i2}$) & -0.002 & -0.109, 0.104 & 94.1 & -0.001 & -0.116, 0.114 & 93.8 & No \\
\midrule[0.3pt]\bottomrule[1pt]
\caption{$^a$The coverage probability is the percent of runs in which the 95\% confidence interval overlapped zero.} \\
\caption{$^b$The robustness test correctly signals outcome-to-outcome interference in the model under the conditions of simulation set 1 or simulation set 2} \\
\caption{Notes: Each simulation consisted of 1,000 runs of 5,000 observations, where each observation represented a sibling pair. Subscripts $i$ and $j$ indicate cluster and sibling, respectively. $T_{ij}$ is the binary treatment, $Y_{ij}$ is the continuous outcome, $U^\prime_{i}$ is the unobserved fixed effect, $C_i$ is the pre-treatment observed fixed effect, and $D_i=Y_{i2}-Y_{i1}$ is the gain-score. The following baseline parameters (linear effects) were fixed across all simulations: $\delta=3.0$ ($T_{ij}\rightarrow Y_{ij}$), $\chi=1.0$ ($U^\prime_{i}\rightarrow T_{i1}$), $\gamma=2.0$ ($U^\prime_{i}\rightarrow T_{i2}$), and $\psi=5.0$ ($U^\prime_{i}\rightarrow Y_{ij}$). The values of $\eta$ ($Y_{i1} \rightarrow Y_{i2}$) and $\pi$ ($C_{i}\hdashrule[0.5ex]{0.6cm}{0.5pt}{2pt} U^\prime_{i}$) varied by simulation set. Within each sample, we conducted the robustness test by regressing $D_i$ on $T_{i1}$, $T_{i2}$, and $C_i$ to assess outcome-to-outcome interference (i.e., whether $\eta$ was non-zero). $b_C$ is the partial regression coefficient for $C_i$. If the robustness test is valid, $b_C=0$ indicates no outcome-to-outcome interference, and $b_C\neq0$ indicates outcome-to-outcome interference. Abbreviations: "CI" confidence interval, "Fig." figure, "SS1" simulation set 1, "SS2" simulation set 2.} \\
\end{longtable}
\end{footnotesize}
\end{onehalfspacing}
\end{landscape}
\clearpage
}

Results from simulations 2.1 and 2.2 indicate that the robustness test was only valid under the models in \textbf{Figures 2B and 2C} if $\pi=0$; if $C_i$ and $U_i^\prime$ were not directly associated, then $C_i$ must have been directly associated with a treatment or directly and equally associated with both outcomes for the robustness test to work. Otherwise, the test was invalid because $C_i$ was not associated with any other variable (\textbf{Figures 1D, 2A, 2D, and 2F}) or because $C_i$ was differentially associated with siblings’ outcomes (\textbf{Figure 2E}).

In a post-hoc analysis, we re-ran all simulation sets under the model in \textbf{Figure 2A} such that $C_i$ was directly associated with only one sibling’s treatment: $\nu=1.5$ ($C_i\hdashrule[0.5ex]{0.6cm}{0.5pt}{2pt}T_{i2}$) and $\tau=0$ ($C_i\hdashrule[0.5ex]{0.6cm}{0.5pt}{2pt}T_{i1}$), or $\nu=0$ and $\tau=0.7$. Results were consistent with main findings, so $C_i$ is a valid observed FE in these settings (see \textbf{Appendix}).


\section{Discussion}

In this paper, we demonstrated a simple robustness test for assessing bias from outcome-to-outcome interference in linear two-sibling FE models. The test involves regressing a gain-score on a pre-treatment observed FE and on both siblings’ treatments. For the test to work, the observed FE must be directly associated with an sibling-invariant confounder on treatments and outcomes, directly associated with at least one sibling’s treatment, or directly and equally associated with both siblings’ outcomes. Further, the robustness test collapses if the observed FE is directly and differentially associated with siblings’ outcomes or if an individual’s outcome affects their sibling’s treatment, although the latter violates the treatment estimation in a sibling FE model regardless.\cite{frisell2012, sjolander2016, frisell2021, mallinson2021} Failure to meet this criterion does not necessarily warrant a variable’s exclusion from a FE regression model. For example, the regression should adjust for an observed cluster-level variable that differentially affects siblings’ treatments and outcomes because it is a confounder, but it is not a valid observed FE for our robustness test.

Finding an observed FE for the robustness test is data dependent. Birth registries are commonplace in sibling FE analyses\cite{khashan2014, khashan2015, hvolgaard2016, axelsson2019, mallinson2020, mallinson2021} and may provide several candidate variables. For instance, United States birth records record maternal demographic information that are fixed and related to family health history, such as nativity and ethnicity.\cite{qin2010, creanga2012, nchs} Other sources may not have a similarly wide breadth of sibling-shared characteristics. Additionally, these shared characteristics may have complex links to siblings’ characteristics, and researchers must interrogate how candidate observed FEs relate the treatments and outcomes. For this reason, we considered only pre-treatment variables as candidate observed FEs for the robustness test. Recent methodological guidance discourages conditioning on post-treatment variables to minimize bias in treatment estimates.\cite{elwert2014, vwbook} In our sibling FE model, conditioning on an observed FE that is causally affected by both siblings’ treatments would induce such bias in the treatment estimate, so it is also possible that it would invalidate the observed FE’s performance in the robustness test. Using an observed FE that is measured prior to siblings’ treatments alleviates this concern. More broadly, we recommend graphical modeling to determine a variable’s usefulness for the robustness test\cite{pearl2013, tennant2021} and, when possible, running multiple robustness tests with valid observed FEs to thoroughly screen possible outcome-to-outcome interference.

We also highlight the flexibility of this robustness test. Despite relying on gain-scores, it is applicable to studies with different FE strategies, such as dummy score regression\cite{kim2021} or conditional maximum likelihood estimation.\cite{sjolander2016} The data may require reshaping such that each observation consists of a dyadic pair. Furthermore, the robustness test may be used in studies of other types of dyadic pairs and not only in sibling comparison analyses. 

There are some limitations to this paper. We focused on settings with linear homogenous effects. Our test may be invalid in models with nonlinear relationships, including those with binary outcomes. Furthermore, we restricted our attention to dyadic sibling models, although this test may work in data with clusters of three or more siblings—for example, one could run the robustness test on randomly selected pairs from sibships. Finally, we omitted models with shared mediator or collider variables, which can induce bias in sibling FE models.\cite{sjolander2017} These factors may also invalidate our robustness test.

In sum, our gain-score robustness test is a useful tool for validating two-sibling FE models. Researchers who conduct sibling comparison studies should take advantage of this test to evaluate potential bias from outcome-to-outcome interference.

\newpage

\newpage

\appendix

\begin{titlepage}
  \centering
  \vspace*{\fill}
  \vskip 60pt
  \LARGE Assessing Outcome-to-Outcome Interference in Sibling Fixed Effects Models \\
 \textit{Appendix} \par
  \vskip 3em
  \large 
  \vspace*{\fill}
\end{titlepage}

\small{}

\renewcommand{\thepage}{S\arabic{page}}

\section{Derivations for Graphical Models} 

\subsection{Baseline model without outcome-to-outcome interference} 

We derive the gain-score regression for the baseline model in \textbf{Figure 1A} to demonstrate how the regression identifies $\delta$ using Wright's path-tracing rules.\cite{pearl2013} The method following applies to settings with variables that are standardized to mean zero and unit variance: between any two variables, the covariance ($\sigma$) is equal to the sum of products of path coefficients along unconditionally open paths that connect the variables. We can use these values to compute partial regression coefficients from linear graphical models. This method is applicable to settings with non-standardized variables with additional computation. For ease of demonstration, we assume models with standardized variables. See Pearl (2013) for additional details.\cite{pearl2013}

Our gain-score regression follows:
\begin{equation*}
D_i=b_1T_{i1}+b_2T_{i2}+e_i.
\end{equation*}

\noindent We want to derive $b_1$ (the partial regression coefficient for $T_{i1}$) and $b_2$ (the partial regression coefficient for $T_{i2}$). To do this, we use the following formulas: 
\begin{flalign*}
b_1 &= \frac{\sigma_{D_{i}T_{i1}}-\sigma_{D_{i}T_{i2}}*\sigma_{T_{i1}T_{i2}}}{1-{\sigma_{T_{i1}T_{i2}}}^2} 
\end{flalign*}
\begin{flalign*}
b_2 &= \frac{\sigma_{D_{i}T_{i2}}-\sigma_{D_{i}T_{i1}}*\sigma_{T_{i1}T_{i2}}}{1-{\sigma_{T_{i1}T_{i2}}}^2} 
\end{flalign*}
Thus, we need to compute $\sigma_{D_{i}T_{i1}}$, $\sigma_{D_{i}T_{i2}}$, and $\sigma_{T_{i1}T_{i2}}$. Following Wright's path-tracing rules, we calculate the following: 
\begin{flalign*}
\sigma_{D_{i}T_{i1}} &= -\psi\chi-\delta+\delta\chi\gamma+\psi\chi \\
&=-\delta+\delta\chi\gamma
\end{flalign*}
\begin{flalign*}
\sigma_{D_{i}T_{i2}} &= -\psi\gamma-\delta\chi\gamma+\delta+\psi\gamma \\
&=-\delta\chi\gamma+\delta
\end{flalign*}
\begin{flalign*}
\sigma_{T_{i1}T_{i2}} &= \chi\gamma
\end{flalign*}
When we compute $b_1$, we identify the negative value of the treatment effect:
\begin{flalign*}
b_1 &= \frac{-\delta+\delta\chi\gamma-(-\delta\chi\gamma+\delta)(\chi\gamma)}{1-{(\chi\gamma)}^2} \\
&= -\delta
\end{flalign*}
Likewise, when we compute $b_2$, we identify the treatment effect precisely:
\begin{flalign*}
b_2 &= \frac{-\delta\chi\gamma+\delta-(-\delta+\delta\chi\gamma)(\chi\gamma)}{1-{(\chi\gamma)}^2} \\
&= \delta
\end{flalign*}

\subsection{Baseline model with outcome-to-outcome interference} 

Following the steps of the prior subsection, we derive the partial regression coefficients $b_1$ and $b_2$ given the model in \textbf{Figure 1B}. Unlike that of \textbf{Figure 1A}, this model includes outcome-to-outcome interference. We compute $\sigma_{D_{i}T_{i1}}$, $\sigma_{D_{i}T_{i2}}$, and $\sigma_{T_{i1}T_{i2}}$ with Wright's path-tracing rules: 
\begin{flalign*}
\sigma_{D_{i}T_{i1}} &= -\psi\chi-\delta+\psi\chi\eta+\delta\eta+\delta\chi\gamma+\psi\chi \\
&=-\delta+\psi\chi\eta+\delta\eta+\delta\chi\gamma
\end{flalign*}
\begin{flalign*}
\sigma_{D_{i}T_{i2}} &= -\psi\gamma-\delta\chi\gamma+\psi\gamma\eta+\delta\chi\gamma\eta+\delta+\psi\gamma \\
&=-\delta\chi\gamma+\psi\gamma\eta+\delta\chi\gamma\eta+\delta
\end{flalign*}
\begin{flalign*}
\sigma_{T_{i1}T_{i2}} &= \chi\gamma
\end{flalign*}
We compute $b_1$, which does not identify (the negative value of) the treatment effect, nor a simple function of it: 
\begin{flalign*}
b_1 &= \frac{-\delta+\psi\chi\eta+\delta\eta+\delta\chi\gamma-(-\delta\chi\gamma+\psi\gamma\eta+\delta\chi\gamma\eta+\delta)(\chi\gamma)}{1-{(\chi\gamma)}^2} \\
&= -\delta+\delta\eta+\frac{\psi\chi\eta(1-{\gamma}^2)}{1-{(\chi\gamma)}^2}
\end{flalign*}
As anticipated, $b_2$ does not identify the treatment effect nor a simple function of it: 
\begin{flalign*}
b_2 &= \frac{-\delta\chi\gamma+\psi\gamma\eta+\delta\chi\gamma\eta+\delta-(\delta+\psi\chi\eta+\delta\eta+\delta\chi\gamma)(\chi\gamma)}{1-{(\chi\gamma)}^2} \\
&= \delta+\frac{\psi\gamma\eta(1-{\chi}^2)}{1-{(\chi\gamma)}^2}
\end{flalign*}

\newpage
\section{Full Simulation Results} 

\textbf{Appendix Table 1} includes additional findings from the main Monte Carlo simulation analyses, specifically the estimated partial regression coefficients and accompanying 95\% confidence intervals for both siblings' treatments. These additional results corroborate the primary findings that we show in the main text. Still, the additional results for the simulations of the treatment-to-outcome interference model (\textbf{Figure 2D}) are noteworthy. In Simulation 1.1 (no outcome-to-outcome interference and the observed FE is related to the unobserved FE), we find that neither $b_1$ nor $b_2$ identify (a simple function of) the focal treatment effect, yet $b_C$ correctly signals the absence of outcome-to-outcome interference. Mallinson and Elwert (2021) demonstrate that we cannot precisely identify treatment effects in this model, but we can identify lower bounds of one of the treatment-to-outcome interference effects (which is more commonly known as a "spillover effect").\cite{mallinson2021}

\afterpage{%
\begin{landscape}
\begin{onehalfspacing}
\begin{footnotesize}
\centering
\setlength\LTleft{0pt}
\setlength\LTright{0pt}
\begin{longtable}{l l S[table-format=1.3] c S[table-format=1.3] c S[table-format=1.3] c S[table-format=2.1]}
\captionsetup{justification=justified, singlelinecheck=false}
\caption{\textbf{Appendix Table 1.} Full Monte Carlo simulation results} \\
\toprule[1pt]\midrule[0.3pt] 
\textbf{Model} & \textbf{Additional Parameters} & \multicolumn{2}{c}{\textbf{Sib. 1's Treatment ($T_{i1}$)}} & \multicolumn{2}{c}{\textbf{Sib. 2's Treatment ($T_{i2}$)}} & \multicolumn{3}{c}{\textbf{Observed Fixed Effect ($C_i$)}} \\ \cline{3-4} \cline{5-6} \cline{7-9} 
\textbf{(Fig.)} & \textbf{(Including Baseline)} & $\widehat{b_1}$ & \textbf{\textit{95\% CI}} & $\widehat{b_2}$ & \textbf{\textit{95\% CI}} & $\widehat{b_C}$ & \textbf{\textit{95\% CI}} & \textbf{\textit{Cov. (\%)}}$^a$ \\
\hline
\rowcolor{lgray2}
\multicolumn{9}{l}{\textbf{Simulation Set 1.1}} \\
\rowcolor{lgray2}
\multicolumn{9}{l}{\textbf{$\eta=0$ ($Y_{i1}\rightarrow Y_{i2}$), $\pi=0.5$ ($C_{i}\hdashrule[0.5ex]{0.6cm}{0.5pt}{2pt} U^\prime_{i}$)}} \\
\hline
\textbf{1D} & None & -2.998 & -3.095, -2.901 & 2.997 & 2.893, 3.102 & -0.000 & -0.105, 0.105 & 95.1 \\
\rowcolor{lightgray}
\textbf{2A} & $\phi=0.9$ ($T_{i1}\rightarrow T_{i2}$) & -2.997 & -3.136, 2.859 & 2.998 & 2.860, 3.136 & -0.000 & -0.105, 0.104 & 95.5 \\
\textbf{2B} & $\tau=0.7$ ($C_{i}\hdashrule[0.5ex]{0.6cm}{0.5pt}{2pt} T_{i1}$), $\nu=1.5$ ($C_{i}\hdashrule[0.5ex]{0.6cm}{0.5pt}{2pt} T_{i2}$) & -3.000 & -3.101, -2.898 & 3.001 & 2.891, 3.111 & -0.001 & -0.116, 0.114 & 94.5 \\
\rowcolor{lightgray}
\textbf{2C} & $\lambda=2.5$ ($C_{i}\hdashrule[0.5ex]{0.6cm}{0.5pt}{2pt} Y_{ij}$) & 2.998 & -3.095, -2.901 & 2.997 & 2.893, 3.102 & -0.000 & -0.105, 0.105 & 95.1 \\
\textbf{2D} & $\theta=1.2$ ($T_{i1}\rightarrow Y_{i2}$), $\kappa=0.6$ ($T_{i2}\rightarrow Y_{i1}$)  & -1.798 & -1.895, -1.701 & 2.397 & 2.293, 2.502 & -0.000 & -0.105, 0.105 & 95.1  \\
\rowcolor{lightgray}
\textbf{2E} & $\lambda=2.5$ ($C_{i}\hdashrule[0.5ex]{0.6cm}{0.5pt}{2pt} Y_{i1}$), $\omega=2.8$ ($C_{i}\hdashrule[0.5ex]{0.6cm}{0.5pt}{2pt} Y_{i2}$) & -2.998 & -3.095, -2.901 & 2.997 & 2.893, 3.102 & 0.300 & 0.195, 0.405 & 0.0  \\
\textbf{2F} & $\mu=0.4$ ($Y_{i1}\rightarrow T_{i2}$) & 2.387 & -2.557, -2.217 & 2.309 & 2.141, 2.477 & 0.033 & -0.071, 0.136 & 89.3 \\
\hline
\rowcolor{lgray2}
\multicolumn{9}{l}{\textbf{Simulation Set 1.2}} \\
\rowcolor{lgray2}
\multicolumn{9}{l}{\textbf{$\eta=0.3$ ($Y_{i1}\rightarrow Y_{i2}$), $\pi=0.5$ ($C_{i}\hdashrule[0.5ex]{0.6cm}{0.5pt}{2pt} U^\prime_{i}$)}} \\
\hline
\textbf{1D} & None & -.514 & -0.611, -0.417 & 4.451 & 4.346, 4.556 & 0.264 & 0.159, 0.370 & 0.3 \\
\rowcolor{lightgray}
\textbf{2A} & $\phi=0.9$ ($T_{i1}\rightarrow T_{i2}$) & -0.845 & -0.989, -0.700 & 4.301 & 4.158, 4.444 & 0.397 & 0.289, 0.505 & 0.0  \\
\textbf{2B} & $\tau=0.7$ ($C_{i}\hdashrule[0.5ex]{0.6cm}{0.5pt}{2pt} T_{i1}$), $\nu=1.5$ ($C_{i}\hdashrule[0.5ex]{0.6cm}{0.5pt}{2pt} T_{i2}$) & -0.522 & -0.625, -0.418 & 4.514 & 4.402, 4.627 & -0.378 & -0.496, -0.260 & 0.0  \\
\rowcolor{lightgray}
\textbf{2C} & $\lambda=2.5$ ($C_{i}\hdashrule[0.5ex]{0.6cm}{0.5pt}{2pt} Y_{ij}$) & -0.514 & -0.611, -0.417 & 4.451 & 4.346, 4.556 & 1.014 & 0.909, 1.120 & 0.0 \\
\textbf{2D} & $\theta=1.2$ ($T_{i1}\rightarrow Y_{i2}$), $\kappa=0.6$ ($T_{i2}\rightarrow Y_{i1}$)  & 0.686 & 0.589,  0.783 & 4.031 & 3.926, 4.136 & 0.264 & 0.159, 0.370 & 0.3  \\
\rowcolor{lightgray}
\textbf{2E} & $\lambda=2.5$ ($C_{i}\hdashrule[0.5ex]{0.6cm}{0.5pt}{2pt} Y_{i1}$), $\omega=2.8$ ($C_{i}\hdashrule[0.5ex]{0.6cm}{0.5pt}{2pt} Y_{i2}$) & -0.514 & -0.611, -0.417 & 4.451 & 4.346, 4.556 & 1.314 & 1.209, 1.420 & 0.0  \\
\textbf{2F} & $\mu=0.4$ ($Y_{i1}\rightarrow T_{i2}$) & -0.448 & -0.630, -0.267 & 3.749 & 3.569, 3.929 & 0.462 & 0.351, 0.573 & 0.0 \\
\hline
\rowcolor{lgray2}
\multicolumn{9}{l}{\textbf{Simulation Set 2.1}} \\
\rowcolor{lgray2}
\multicolumn{9}{l}{\textbf{$\eta=0$ ($Y_{i1}\rightarrow Y_{i2}$), $\pi=0$ ($C_{i}\hdashrule[0.5ex]{0.6cm}{0.5pt}{2pt} U^\prime_{i}$)}} \\
\hline
\textbf{1D} & None & -2.998 & -3.094, -2.901 & 2.997 & 2.894, 3.101 & -0.002 & -0.109, 0.105 & 94.0  \\
\rowcolor{lightgray}
\textbf{2A} & $\phi=0.9$ ($T_{i1}\rightarrow T_{i2}$) & -2.997 & -3.136, -2.859 & 2.998 & 2.861, 3.135 & -0.002 & -0.109, 0.105 & 94.1  \\
\textbf{2B} & $\tau=0.7$ ($C_{i}\hdashrule[0.5ex]{0.6cm}{0.5pt}{2pt} T_{i1}$), $\nu=1.5$ ($C_{i}\hdashrule[0.5ex]{0.6cm}{0.5pt}{2pt} T_{i2}$) &  -3.000 & -3.100, -2.901 & 3.001 & 2.899, 3.103 & -0.002 & -0.112, 0.108 & 94.2  \\
\rowcolor{lightgray}
\textbf{2C} & $\lambda=2.5$ ($C_{i}\hdashrule[0.5ex]{0.6cm}{0.5pt}{2pt} Y_{ij}$) & -2.998 & -3.094, -2.901 & 2.997 & 2.894, 3.101 & -0.002 & -0.109, 0.105 & 94.0 \\
\textbf{2D} & $\theta=1.2$ ($T_{i1}\rightarrow Y_{i2}$), $\kappa=0.6$ ($T_{i2}\rightarrow Y_{i1}$)  & -1.798 & -1.894, -1.701 & 2.397 & 2.294, 2.501 & -0.002 & -0.109, 0.105 & 94.0 \\
\rowcolor{lightgray}
\textbf{2E} & $\lambda=2.5$ ($C_{i}\hdashrule[0.5ex]{0.6cm}{0.5pt}{2pt} Y_{i1}$), $\omega=2.8$ ($C_{i}\hdashrule[0.5ex]{0.6cm}{0.5pt}{2pt} Y_{i2}$) & -2.998 & -3.094, -2.901 & 2.997 & 2.894, 3.101 & 0.298 & 0.191, 0.405 & 0.1  \\
\textbf{2F} & $\mu=0.4$ ($Y_{i1}\rightarrow T_{i2}$) & -2.385 & -2.554, -2.215 & 2.314 & 2.146, 2.481 & -0.002 & -0.109, 0.104 & 94.1 \\
\hline
\rowcolor{lgray2}
\multicolumn{9}{l}{\textbf{Simulation Set 2.2}} \\
\rowcolor{lgray2}
\multicolumn{9}{l}{\textbf{$\eta=0.3$ ($Y_{i1}\rightarrow Y_{i2}$), $\pi=0$ ($C_{i}\hdashrule[0.5ex]{0.6cm}{0.5pt}{2pt} U^\prime_{i}$)}} \\
\hline
\textbf{1D} & None & -0.488 & -0.584, -0.391 & 4.493 & 4.390, 4.597 & -0.002 & -0.109, 0.106 & 93.1 \\
\rowcolor{lightgray}
\textbf{2A} & $\phi=0.9$ ($T_{i1}\rightarrow T_{i2}$) & -0.816 & -0.961, -0.671 & 4.355 & 4.212, 4.498 & -0.001 & -0.113, 0.111 & 93.1  \\
\textbf{2B} & $\tau=0.7$ ($C_{i}\hdashrule[0.5ex]{0.6cm}{0.5pt}{2pt} T_{i1}$), $\nu=1.5$ ($C_{i}\hdashrule[0.5ex]{0.6cm}{0.5pt}{2pt} T_{i2}$) & -0.523 & -0.623, -0.422 & 4.547 & 4.444, 4.650 & -0.825 & -0.936, -0.714 & 0.0  \\
\rowcolor{lightgray}
\textbf{2C} & $\lambda=2.5$ ($C_{i}\hdashrule[0.5ex]{0.6cm}{0.5pt}{2pt} Y_{ij}$) & -0.488 & -0.584, -0.391 & 4.493 & 4.390, 4.597 & 0.749 & 0.641, 0.856 & 0.0 \\
\textbf{2D} & $\theta=1.2$ ($T_{i1}\rightarrow Y_{i2}$), $\kappa=0.6$ ($T_{i2}\rightarrow Y_{i1}$)  & 0.712 & 0.616, 0.809 & 4.073 & 3.970, 4.177 & -0.001 & -0.109, 0.106 & 93.1  \\
\rowcolor{lightgray}
\textbf{2E} & $\lambda=2.5$ ($C_{i}\hdashrule[0.5ex]{0.6cm}{0.5pt}{2pt} Y_{i1}$), $\omega=2.8$ ($C_{i}\hdashrule[0.5ex]{0.6cm}{0.5pt}{2pt} Y_{i2}$) & -0.488 & -0.584, -0.391 & 4.493 & 4.390, 4.597 & 1.049 & 0.941, 1.156 & 0.0  \\
\textbf{2F} & $\mu=0.4$ ($Y_{i1}\rightarrow T_{i2}$) & -0.416 & -0.598, -0.233 & 3.808 & 3.628, 3.988 & -0.001 & -0.116, 0.114 & 93.8 \\
\midrule[0.3pt]\bottomrule[1pt]
\caption{$^a$The coverage probability is the percent of runs in which the 95\% confidence interval overlapped zero.} \\~\\
\caption{Notes: Each simulation consisted of 1,000 runs of 5,000 observations, where each observation represented a sibling pair. Subscripts $i$ and $j$ indicate cluster and sibling, respectively. $T_{ij}$ is the binary treatment, $Y_{ij}$ is the continuous outcome, $U^\prime_{i}$ is the unobserved vector of family-level confounders, $C_i$ is the pre-treatment observed fixed effect, and $D_i=Y_{i2}-Y_{i1}$ is the gain-score. The following baseline parameters (linear effects) were fixed across all simulations: $\delta=3.0$ ($T_{ij}\rightarrow Y_{ij}$), $\chi=1.0$ ($U^\prime_{i}\rightarrow T_{i1}$), $\gamma=2.0$ ($U^\prime_{i}\rightarrow T_{i2}$), and $\psi=5.0$ ($U^\prime_{i}\rightarrow Y_{ij}$). The values of $\eta$ ($Y_{i1} \rightarrow Y_{i2}$) and $\pi$ ($C_{i}\hdashrule[0.5ex]{0.6cm}{0.5pt}{2pt} U^\prime_{i}$) varied by simulation set. Within each sample, we conducted the robustness test by regressing $D_i$ on $T_{i1}$, $T_{i2}$, and $C_i$ to assess outcome-to-outcome interference (i.e., whether $\eta$ was non-zero). $b_C$ is the partial regression coefficient for $C_i$. If the robustness test is valid, $b_C=0$ indicates no outcome-to-outcome interference, and $b_C\neq0$ indicates outcome-to-outcome interference. Abbreviations: "CI" confidence interval, "Cov." coverage probability, "Fig." figure.} \\
\end{longtable}
\end{footnotesize}
\end{onehalfspacing}
\end{landscape}
\clearpage
}

\newpage

\section{Post-Hoc Simulation} 

\textbf{Appendix Table 2} displays results from a post-hoc Monte Carlo simulation analysis. We re-ran all simulation sets as described in the main text for the model in \textbf{Figure 2B}, where the observed FE $C_i$ has a direct association with both siblings' treatments. However, we varied the simulations such that the observed FE had a direct association with only one sibling's treatment—that is, $\nu=1.5$ ($C_i\hdashrule[0.5ex]{0.6cm}{0.5pt}{2pt} T_{i2}$) and $\tau=0$ ($C_i\hdashrule[0.5ex]{0.6cm}{0.5pt}{2pt} T_{i1}$), or $\nu=0$ and $\tau=0.7$. These results were consistent with those of the main analysis, indicating that $C_i$ could be a valid observed FE even if it was directly associated with a single sibling's treatment and with no other variable in the data generating process. 

\afterpage{%
\begin{landscape}
\begin{onehalfspacing}
\begin{footnotesize}
\centering
\setlength\LTleft{0pt}
\setlength\LTright{0pt}
\begin{longtable}{l l S[table-format=1.3] c S[table-format=1.3] c S[table-format=1.3] c S[table-format=2.1]}
\captionsetup{justification=justified, singlelinecheck=false}
\caption{\textbf{Appendix Table 2.} Post-hoc Monte Carlo simulation results} \\
\toprule[1pt]\midrule[0.3pt] 
\multicolumn{2}{l}{\textbf{Additional Parameters}} & & & & & & &  \\
\multicolumn{2}{l}{\textbf{(Including Baseline)}} & \multicolumn{2}{c}{\textbf{Sibling 1's Treatment ($T_{i1}$)}} & \multicolumn{2}{c}{\textbf{Sibling 2's Treatment ($T_{i2}$)}} & \multicolumn{3}{c}{\textbf{Observed Fixed Effect ($C_i$)}} \\ \cline{1-2} \cline{3-4} \cline{5-6} \cline{7-9} 
\textbf{$\tau$ ($C_{i}\hdashrule[0.5ex]{0.6cm}{0.5pt}{2pt} T_{i1}$)} & \textbf{$\nu$ ($C_{i}\hdashrule[0.5ex]{0.6cm}{0.5pt}{2pt} T_{i2}$)} & $\widehat{b_1}$ & \textbf{\textit{95\% CI}} & $\widehat{b_2}$ & \textbf{\textit{95\% CI}} & $\widehat{b_C}$ & \textbf{\textit{95\% CI}} & \textbf{\textit{Cov. (\%)}}$^a$ \\
\hline
\rowcolor{lgray2}
\multicolumn{9}{l}{\textbf{Simulation Set 1.1}} \\
\rowcolor{lgray2}
\multicolumn{9}{l}{\textbf{$\eta=0$ ($Y_{i1}\rightarrow Y_{i2}$), $\pi=0.5$ ($C_{i}\hdashrule[0.5ex]{0.6cm}{0.5pt}{2pt} U^\prime_{i}$)}} \\
\hline
0 & 1.5 & -2.999 & -3.102, -2.897 & 3.000 & 2.884, 3.117 & -0.001 & -0.116, 0.114 & 94.0 \\
\rowcolor{lightgray}
0.7 & 0 & 2.998 & -3.097, -2.900 & 2.998 & 2.896, 3.099 & -0.000 & -0.108, 0.108 & 95.4 \\
\hline
\rowcolor{lgray2}
\multicolumn{9}{l}{\textbf{Simulation Set 1.2}} \\
\rowcolor{lgray2}
\multicolumn{9}{l}{\textbf{$\eta=0.3$ ($Y_{i1}\rightarrow Y_{i2}$), $\pi=0.5$ ($C_{i}\hdashrule[0.5ex]{0.6cm}{0.5pt}{2pt} U^\prime_{i}$)}} \\
\hline
0 & 1.5 & -0.493 & -0.598, -0.388 & 4.335 & 4.216, 4.454 & -0.071 & -0.189, 0.046 & 76.9 \\
\rowcolor{lightgray}
0.7 & 0 & -0.533 & -0.632, -0.434 & 4.572 & 4.470, 4.674 & -0.024 & -0.085, 0.132 & 93.3 \\
\hline
\rowcolor{lgray2}
\multicolumn{9}{l}{\textbf{Simulation Set 2.1}} \\
\rowcolor{lgray2}
\multicolumn{9}{l}{\textbf{$\eta=0$ ($Y_{i1}\rightarrow Y_{i2}$), $\pi=0$ ($C_{i}\hdashrule[0.5ex]{0.6cm}{0.5pt}{2pt} U^\prime_{i}$)}} \\
\hline
0 & 1.5 & -2.999 & -3.102, 2.897 & 3.000 & 2.892, 3.109 & -0.002 & -0.114, 0.110 & 94.2 \\
\rowcolor{lightgray}
0.7 & 0 & 2.999 & -3.096, -2.902 & 2.998 & 2.897, 3.099 & -0.002 & -0.112, 0.108 & 94.1 \\
\hline
\rowcolor{lgray2}
\multicolumn{9}{l}{\textbf{Simulation Set 2.2}} \\
\rowcolor{lgray2}
\multicolumn{9}{l}{\textbf{$\eta=0.3$ ($Y_{i1}\rightarrow Y_{i2}$), $\pi=0$ ($C_{i}\hdashrule[0.5ex]{0.6cm}{0.5pt}{2pt} U^\prime_{i}$)}} \\
\hline
0 & 1.5 & -0.544 & -0.648, -0.439 & 4.420 & 4.310, 4.531 & -0.414 & -0.528, -0.300 & 0.0 \\
\rowcolor{lightgray}
0.7 & 0 & -0.481 & -0.578, -0.384 & 4.578 & 4.477, 4.678 & -0.385 & -0.495, -0.276 & 0.0 \\
\midrule[0.3pt]\bottomrule[1pt]
\caption{$^a$The coverage probability is the percent of runs in which the 95\% confidence interval overlapped zero.} \\
\caption{Notes: Each simulation consisted of 1,000 runs of 5,000 observations, where each observation represented a sibling pair. Subscripts $i$ and $j$ indicate cluster and sibling, respectively. $T_{ij}$ is the binary treatment, $Y_{ij}$ is the continuous outcome, $U^\prime_{i}$ is the unobserved vector of family-level confounders, $C_i$ is the pre-treatment observed fixed effect, and $D_i=Y_{i2}-Y_{i1}$ is the gain-score. The following baseline parameters (linear effects) were fixed across all simulations: $\delta=3.0$ ($T_{ij}\rightarrow Y_{ij}$), $\chi=1.0$ ($U^\prime_{i}\rightarrow T_{i1}$), $\gamma=2.0$ ($U^\prime_{i}\rightarrow T_{i2}$), and $\psi=5.0$ ($U^\prime_{i}\rightarrow Y_{ij}$). The values of $\eta$ ($Y_{i1} \rightarrow Y_{i2}$) and $\pi$ ($C_{i}\hdashrule[0.5ex]{0.6cm}{0.5pt}{2pt} U^\prime_{i}$) varied by simulation set. Within each sample, we conducted the robustness test by regressing $D_i$ on $T_{i1}$, $T_{i2}$, and $C_i$ to assess outcome-to-outcome interference (i.e., whether $\eta$ was non-zero). $b_C$ is the partial regression coefficient for $C_i$. If the robustness test is valid, $b_C=0$ indicates no outcome-to-outcome interference, and $b_C\neq0$ indicates outcome-to-outcome interference. Abbreviations: "CI" confidence interval, "Cov." coverage probability, "Fig." figure.} \\
\end{longtable}
\end{footnotesize}
\end{onehalfspacing}
\end{landscape}
\clearpage
}

\newpage

\section{Main Simulation Code}

We conducted Monte Carlo simulations in Stata Statistical Software: Release 16.\cite{statasoft} 

\begin{singlespace}
\lstinputlisting[language=Stata,style=stata-editor]{sibling_fe_robustness_mainsim_code.do}
\end{singlespace}

\newpage

\section{Post-Hoc Simulation Code}

We conducted Monte Carlo simulations in Stata Statistical Software: Release 16.\cite{statasoft} 

\begin{singlespace}
\begin{lstlisting}[language=Stata,style=stata-editor]
/*
Assessing Outcome-to-Outcome Interference in Sibling Fixed Effects Models

Post-Hoc Simulation Code

David C. Mallinson

September 27, 2021
*/

/************************************/
*Preliminaries
/************************************/

/*
Defintions:

t_1 = sibling 1's treatment
t_2 = sibling 2's treatment
y_1 = sibling 1's outcome
y_2 = sibling 2's outcome
u = unobserved fixed effect
c = observed family-level characteristic
d = gain-score (y_2-y_1)
*/


/************************************/
*Model 2B, Sim 1.1
*No outcome-to-outcome spillover
*C related to U
*C unrelated to T1
/************************************/

**clear previous data
clear all
set more off

**monte carlo simulation settings 
global nobs=5000 /*observations per run*/
global nmc=1000 /*number of runs*/
set seed 54385 /*seed for random number generator -- can change*/
set obs $nobs /*sets observations within each run*/

**set parameters (causal effects in models)
scalar delta=3 /*t_1 on y_1, t_2 on y_2*/

scalar psi=5 /*u on y_1, u on y_2*/
scalar chi=1 /*u on t_1*/
scalar gamma=2 /*u on t_2*/
scalar pi=0.5 /*u on c*/

scalar eta=0 /*y_1 on y_2*/

scalar phi=0.9 /*t_1 on t_2*/
scalar tau=0 /*c on t_1*/
scalar nu=1.5 /*c on t_2*/
scalar lambda=2.5 /*c on y_1, c on y_2*/
scalar omega=2.8 /*c on y_2*/
scalar theta=1.2 /*t_1 on y_2*/
scalar kappa=0.6 /*t_2 on y_1*/
scalar mu=0.4 /*y_1 on t_2*/

**generate variables

gen u=. /*unobserved fixed effect*/
gen c=. /*observed family-level characteristic*/ 
gen t_1=. /*sibling 1's treatment*/
gen t_2=. /*sibling 2's treatment*/
gen y_1=. /*sibling 1's outcome*/
gen y_2=. /*sibling 2's outcome*/
gen d=. /*gain-score (y_2-y_1)*/

**simulation 

tempname sim
/*
for t1, t2, and c:
X_mc is the regression coefficient
X_ub_mc is the 95% confidence interval upper bound
X_lb_mc is the 95% confidence interval lower bound
*/
postfile `sim' t1_mc t1_ub_mc t1_lb_mc ///
	t2_mc t2_ub_mc t2_lb_mc ///
	r_mc r_ub_mc r_lb_mc using sim, replace
quietly {

forvalues i=1/$nmc {

	/*note: for c, y_1, and y_2 equations, "rnormal(0,1)" is residual*/

	replace u=rnormal(0,1)
	replace c=(pi*u)+rnormal(0,1)>1
	replace t_1=(chi*u)+(tau*c)>-0.2
	replace t_2=(gamma*u)+(nu*c)>1
	replace y_1=(delta*t_1)+(psi*u)+rnormal(0,1)
	replace y_2=(delta*t_2)+(psi*u)+(eta*y_1)+rnormal(0,1)
	replace d=y_2-y_1
	
	/*gain-score regression*/
	reg d t_1 t_2 c
	
	/*store regression output*/
	*sibling 1 treatment
	scalar t1_mc=_b[t_1]
	scalar t1_ub_mc=(_b[t_1]+(1.96*_se[t_1])) /*95% CI upper bound*/
	scalar t1_lb_mc=(_b[t_1]-(1.96*_se[t_1])) /*95% CI lower bound*/
	*sibling 2 treatment
	scalar t2_mc=_b[t_2]
	scalar t2_ub_mc=(_b[t_2]+(1.96*_se[t_2])) /*95% CI upper bound*/
	scalar t2_lb_mc=(_b[t_2]-(1.96*_se[t_2])) /*95% CI lower bound*/
	*robustness test result
	scalar r_mc=_b[c] 
	scalar r_ub_mc=(_b[c]+(1.96*_se[c])) /*95% CI upper bound*/
	scalar r_lb_mc=(_b[c]-(1.96*_se[c])) /*95% CI lower bound*/
	
	post `sim' (t1_mc) (t1_ub_mc) (t1_lb_mc) ///
		(t2_mc) (t2_ub_mc) (t2_lb_mc) ///
		(r_mc) (r_ub_mc) (r_lb_mc)
}
}
postclose `sim'

use sim, clear
summarize

**store mean of results, save to append with other simulation results
egen t1=mean(t1_mc)
egen t1_ub=mean(t1_ub_mc)
egen t1_lb=mean(t1_lb_mc)
egen t2=mean(t2_mc)
egen t2_ub=mean(t2_ub_mc)
egen t2_lb=mean(t2_lb_mc)
egen r=mean(r_mc)
egen r_ub=mean(r_ub_mc)
egen r_lb=mean(r_lb_mc)
gen id=1 /*indicates simulation order*/

**generate coverage probability (% of confidence intervals that overlap zero)
gen cov_pre=0
replace cov_pre=1 if r_lb_mc<0 & 0<r_ub_mc
tab cov_pre
egen cov=mean(cov_pre)

keep t1 t1_ub t1_lb t2 t2_ub t2_lb r r_ub r_lb cov id
duplicates drop
duplicates report 

**save data 
save "ph_sim1.dta", replace


/************************************/
*Model 2B, Sim 1.2
*With outcome-to-outcome spillover
*C related to U
*C unrelated to T1
/************************************/

**clear previous data
clear all
set more off

**monte carlo simulation settings 
global nobs=5000 /*observations per run*/
global nmc=1000 /*number of runs*/
set seed 54385 /*seed for random number generator -- can change*/
set obs $nobs /*sets observations within each run*/

**set parameters (causal effects in models)
scalar delta=3 /*t_1 on y_1, t_2 on y_2*/

scalar psi=5 /*u on y_1, u on y_2*/
scalar chi=1 /*u on t_1*/
scalar gamma=2 /*u on t_2*/
scalar pi=0.5 /*u on c*/

scalar eta=0.3 /*y_1 on y_2*/

scalar phi=0.9 /*t_1 on t_2*/
scalar tau=0 /*c on t_1*/
scalar nu=1.5 /*c on t_2*/
scalar lambda=2.5 /*c on y_1, c on y_2*/
scalar omega=2.8 /*c on y_2*/
scalar theta=1.2 /*t_1 on y_2*/
scalar kappa=0.6 /*t_2 on y_1*/
scalar mu=0.4 /*y_1 on t_2*/

**generate variables

gen u=. /*unobserved fixed effect*/
gen c=. /*observed family-level characteristic*/ 
gen t_1=. /*sibling 1's treatment*/
gen t_2=. /*sibling 2's treatment*/
gen y_1=. /*sibling 1's outcome*/
gen y_2=. /*sibling 2's outcome*/
gen d=. /*gain-score (y_2-y_1)*/

**simulation 

tempname sim
/*
for t1, t2, and c:
X_mc is the regression coefficient
X_ub_mc is the 95% confidence interval upper bound
X_lb_mc is the 95% confidence interval lower bound
*/
postfile `sim' t1_mc t1_ub_mc t1_lb_mc ///
	t2_mc t2_ub_mc t2_lb_mc ///
	r_mc r_ub_mc r_lb_mc using sim, replace
quietly {

forvalues i=1/$nmc {

	/*note: for c, y_1, and y_2 equations, "rnormal(0,1)" is residual*/

	replace u=rnormal(0,1)
	replace c=(pi*u)+rnormal(0,1)>1
	replace t_1=(chi*u)+(tau*c)>-0.2
	replace t_2=(gamma*u)+(nu*c)>1
	replace y_1=(delta*t_1)+(psi*u)+rnormal(0,1)
	replace y_2=(delta*t_2)+(psi*u)+(eta*y_1)+rnormal(0,1)
	replace d=y_2-y_1
	
	/*gain-score regression*/
	reg d t_1 t_2 c
	
	/*store regression output*/
	*sibling 1 treatment
	scalar t1_mc=_b[t_1]
	scalar t1_ub_mc=(_b[t_1]+(1.96*_se[t_1])) /*95% CI upper bound*/
	scalar t1_lb_mc=(_b[t_1]-(1.96*_se[t_1])) /*95% CI lower bound*/
	*sibling 2 treatment
	scalar t2_mc=_b[t_2]
	scalar t2_ub_mc=(_b[t_2]+(1.96*_se[t_2])) /*95% CI upper bound*/
	scalar t2_lb_mc=(_b[t_2]-(1.96*_se[t_2])) /*95% CI lower bound*/
	*robustness test result
	scalar r_mc=_b[c] 
	scalar r_ub_mc=(_b[c]+(1.96*_se[c])) /*95% CI upper bound*/
	scalar r_lb_mc=(_b[c]-(1.96*_se[c])) /*95% CI lower bound*/
	
	post `sim' (t1_mc) (t1_ub_mc) (t1_lb_mc) ///
		(t2_mc) (t2_ub_mc) (t2_lb_mc) ///
		(r_mc) (r_ub_mc) (r_lb_mc)
}
}
postclose `sim'

use sim, clear
summarize

**store mean of results, save to append with other simulation results
egen t1=mean(t1_mc)
egen t1_ub=mean(t1_ub_mc)
egen t1_lb=mean(t1_lb_mc)
egen t2=mean(t2_mc)
egen t2_ub=mean(t2_ub_mc)
egen t2_lb=mean(t2_lb_mc)
egen r=mean(r_mc)
egen r_ub=mean(r_ub_mc)
egen r_lb=mean(r_lb_mc)
gen id=2 /*indicates simulation order*/

**generate coverage probability (% of confidence intervals that overlap zero)
gen cov_pre=0
replace cov_pre=1 if r_lb_mc<0 & 0<r_ub_mc
tab cov_pre
egen cov=mean(cov_pre)

keep t1 t1_ub t1_lb t2 t2_ub t2_lb r r_ub r_lb cov id
duplicates drop
duplicates report 

**save data 
save "ph_sim2.dta", replace


/************************************/
*Model 2B, Sim 2.1
*No outcome-to-outcome spillover
*C unrelated to U
*C unrelated to T1
/************************************/

**clear previous data
clear all
set more off

**monte carlo simulation settings 
global nobs=5000 /*observations per run*/
global nmc=1000 /*number of runs*/
set seed 54385 /*seed for random number generator -- can change*/
set obs $nobs /*sets observations within each run*/

**set parameters (causal effects in models)
scalar delta=3 /*t_1 on y_1, t_2 on y_2*/

scalar psi=5 /*u on y_1, u on y_2*/
scalar chi=1 /*u on t_1*/
scalar gamma=2 /*u on t_2*/
scalar pi=0 /*u on c*/

scalar eta=0 /*y_1 on y_2*/

scalar phi=0.9 /*t_1 on t_2*/
scalar tau=0 /*c on t_1*/
scalar nu=1.5 /*c on t_2*/
scalar lambda=2.5 /*c on y_1, c on y_2*/
scalar omega=2.8 /*c on y_2*/
scalar theta=1.2 /*t_1 on y_2*/
scalar kappa=0.6 /*t_2 on y_1*/
scalar mu=0.4 /*y_1 on t_2*/

**generate variables

gen u=. /*unobserved fixed effect*/
gen c=. /*observed family-level characteristic*/ 
gen t_1=. /*sibling 1's treatment*/
gen t_2=. /*sibling 2's treatment*/
gen y_1=. /*sibling 1's outcome*/
gen y_2=. /*sibling 2's outcome*/
gen d=. /*gain-score (y_2-y_1)*/

**simulation 

tempname sim
/*
for t1, t2, and c:
X_mc is the regression coefficient
X_ub_mc is the 95% confidence interval upper bound
X_lb_mc is the 95% confidence interval lower bound
*/
postfile `sim' t1_mc t1_ub_mc t1_lb_mc ///
	t2_mc t2_ub_mc t2_lb_mc ///
	r_mc r_ub_mc r_lb_mc using sim, replace
quietly {

forvalues i=1/$nmc {

	/*note: for c, y_1, and y_2 equations, "rnormal(0,1)" is residual*/

	replace u=rnormal(0,1)
	replace c=(pi*u)+rnormal(0,1)>1
	replace t_1=(chi*u)+(tau*c)>-0.2
	replace t_2=(gamma*u)+(nu*c)>1
	replace y_1=(delta*t_1)+(psi*u)+rnormal(0,1)
	replace y_2=(delta*t_2)+(psi*u)+(eta*y_1)+rnormal(0,1)
	replace d=y_2-y_1
	
	/*gain-score regression*/
	reg d t_1 t_2 c
	
	/*store regression output*/
	*sibling 1 treatment
	scalar t1_mc=_b[t_1]
	scalar t1_ub_mc=(_b[t_1]+(1.96*_se[t_1])) /*95% CI upper bound*/
	scalar t1_lb_mc=(_b[t_1]-(1.96*_se[t_1])) /*95% CI lower bound*/
	*sibling 2 treatment
	scalar t2_mc=_b[t_2]
	scalar t2_ub_mc=(_b[t_2]+(1.96*_se[t_2])) /*95% CI upper bound*/
	scalar t2_lb_mc=(_b[t_2]-(1.96*_se[t_2])) /*95% CI lower bound*/
	*robustness test result
	scalar r_mc=_b[c] 
	scalar r_ub_mc=(_b[c]+(1.96*_se[c])) /*95% CI upper bound*/
	scalar r_lb_mc=(_b[c]-(1.96*_se[c])) /*95% CI lower bound*/
	
	post `sim' (t1_mc) (t1_ub_mc) (t1_lb_mc) ///
		(t2_mc) (t2_ub_mc) (t2_lb_mc) ///
		(r_mc) (r_ub_mc) (r_lb_mc)
}
}
postclose `sim'

use sim, clear
summarize

**store mean of results, save to append with other simulation results
egen t1=mean(t1_mc)
egen t1_ub=mean(t1_ub_mc)
egen t1_lb=mean(t1_lb_mc)
egen t2=mean(t2_mc)
egen t2_ub=mean(t2_ub_mc)
egen t2_lb=mean(t2_lb_mc)
egen r=mean(r_mc)
egen r_ub=mean(r_ub_mc)
egen r_lb=mean(r_lb_mc)
gen id=3 /*indicates simulation order*/

**generate coverage probability (% of confidence intervals that overlap zero)
gen cov_pre=0
replace cov_pre=1 if r_lb_mc<0 & 0<r_ub_mc
tab cov_pre
egen cov=mean(cov_pre)

keep t1 t1_ub t1_lb t2 t2_ub t2_lb r r_ub r_lb cov id
duplicates drop
duplicates report 

**save data 
save "ph_sim3.dta", replace


/************************************/
*Model 2B, Sim 2.2
*With outcome-to-outcome spillover
*C unrelated to U
*C unrelated to T1
/************************************/

**clear previous data
clear all
set more off

**monte carlo simulation settings 
global nobs=5000 /*observations per run*/
global nmc=1000 /*number of runs*/
set seed 54385 /*seed for random number generator -- can change*/
set obs $nobs /*sets observations within each run*/

**set parameters (causal effects in models)
scalar delta=3 /*t_1 on y_1, t_2 on y_2*/

scalar psi=5 /*u on y_1, u on y_2*/
scalar chi=1 /*u on t_1*/
scalar gamma=2 /*u on t_2*/
scalar pi=0 /*u on c*/

scalar eta=0.3 /*y_1 on y_2*/

scalar phi=0.9 /*t_1 on t_2*/
scalar tau=0 /*c on t_1*/
scalar nu=1.5 /*c on t_2*/
scalar lambda=2.5 /*c on y_1, c on y_2*/
scalar omega=2.8 /*c on y_2*/
scalar theta=1.2 /*t_1 on y_2*/
scalar kappa=0.6 /*t_2 on y_1*/
scalar mu=0.4 /*y_1 on t_2*/

**generate variables

gen u=. /*unobserved fixed effect*/
gen c=. /*observed family-level characteristic*/ 
gen t_1=. /*sibling 1's treatment*/
gen t_2=. /*sibling 2's treatment*/
gen y_1=. /*sibling 1's outcome*/
gen y_2=. /*sibling 2's outcome*/
gen d=. /*gain-score (y_2-y_1)*/

**simulation 

tempname sim
/*
for t1, t2, and c:
X_mc is the regression coefficient
X_ub_mc is the 95% confidence interval upper bound
X_lb_mc is the 95% confidence interval lower bound
*/
postfile `sim' t1_mc t1_ub_mc t1_lb_mc ///
	t2_mc t2_ub_mc t2_lb_mc ///
	r_mc r_ub_mc r_lb_mc using sim, replace
quietly {

forvalues i=1/$nmc {

	/*note: for c, y_1, and y_2 equations, "rnormal(0,1)" is residual*/

	replace u=rnormal(0,1)
	replace c=(pi*u)+rnormal(0,1)>1
	replace t_1=(chi*u)+(tau*c)>-0.2
	replace t_2=(gamma*u)+(nu*c)>1
	replace y_1=(delta*t_1)+(psi*u)+rnormal(0,1)
	replace y_2=(delta*t_2)+(psi*u)+(eta*y_1)+rnormal(0,1)
	replace d=y_2-y_1
	
	/*gain-score regression*/
	reg d t_1 t_2 c
	
	/*store regression output*/
	*sibling 1 treatment
	scalar t1_mc=_b[t_1]
	scalar t1_ub_mc=(_b[t_1]+(1.96*_se[t_1])) /*95% CI upper bound*/
	scalar t1_lb_mc=(_b[t_1]-(1.96*_se[t_1])) /*95% CI lower bound*/
	*sibling 2 treatment
	scalar t2_mc=_b[t_2]
	scalar t2_ub_mc=(_b[t_2]+(1.96*_se[t_2])) /*95% CI upper bound*/
	scalar t2_lb_mc=(_b[t_2]-(1.96*_se[t_2])) /*95% CI lower bound*/
	*robustness test result
	scalar r_mc=_b[c] 
	scalar r_ub_mc=(_b[c]+(1.96*_se[c])) /*95% CI upper bound*/
	scalar r_lb_mc=(_b[c]-(1.96*_se[c])) /*95% CI lower bound*/
	
	post `sim' (t1_mc) (t1_ub_mc) (t1_lb_mc) ///
		(t2_mc) (t2_ub_mc) (t2_lb_mc) ///
		(r_mc) (r_ub_mc) (r_lb_mc)
}
}
postclose `sim'

use sim, clear
summarize

**store mean of results, save to append with other simulation results
egen t1=mean(t1_mc)
egen t1_ub=mean(t1_ub_mc)
egen t1_lb=mean(t1_lb_mc)
egen t2=mean(t2_mc)
egen t2_ub=mean(t2_ub_mc)
egen t2_lb=mean(t2_lb_mc)
egen r=mean(r_mc)
egen r_ub=mean(r_ub_mc)
egen r_lb=mean(r_lb_mc)
gen id=4 /*indicates simulation order*/

**generate coverage probability (% of confidence intervals that overlap zero)
gen cov_pre=0
replace cov_pre=1 if r_lb_mc<0 & 0<r_ub_mc
tab cov_pre
egen cov=mean(cov_pre)

keep t1 t1_ub t1_lb t2 t2_ub t2_lb r r_ub r_lb cov id
duplicates drop
duplicates report 

**save data 
save "ph_sim4.dta", replace


/************************************/
*Model 2B, Sim 1.1
*No outcome-to-outcome spillover
*C related to U
*C unrelated to T1
/************************************/

**clear previous data
clear all
set more off

**monte carlo simulation settings 
global nobs=5000 /*observations per run*/
global nmc=1000 /*number of runs*/
set seed 54385 /*seed for random number generator -- can change*/
set obs $nobs /*sets observations within each run*/

**set parameters (causal effects in models)
scalar delta=3 /*t_1 on y_1, t_2 on y_2*/

scalar psi=5 /*u on y_1, u on y_2*/
scalar chi=1 /*u on t_1*/
scalar gamma=2 /*u on t_2*/
scalar pi=0.5 /*u on c*/

scalar eta=0 /*y_1 on y_2*/

scalar phi=0.9 /*t_1 on t_2*/
scalar tau=0.7 /*c on t_1*/
scalar nu=0 /*c on t_2*/
scalar lambda=2.5 /*c on y_1, c on y_2*/
scalar omega=2.8 /*c on y_2*/
scalar theta=1.2 /*t_1 on y_2*/
scalar kappa=0.6 /*t_2 on y_1*/
scalar mu=0.4 /*y_1 on t_2*/

**generate variables

gen u=. /*unobserved fixed effect*/
gen c=. /*observed family-level characteristic*/ 
gen t_1=. /*sibling 1's treatment*/
gen t_2=. /*sibling 2's treatment*/
gen y_1=. /*sibling 1's outcome*/
gen y_2=. /*sibling 2's outcome*/
gen d=. /*gain-score (y_2-y_1)*/

**simulation 

tempname sim
/*
for t1, t2, and c:
X_mc is the regression coefficient
X_ub_mc is the 95% confidence interval upper bound
X_lb_mc is the 95% confidence interval lower bound
*/
postfile `sim' t1_mc t1_ub_mc t1_lb_mc ///
	t2_mc t2_ub_mc t2_lb_mc ///
	r_mc r_ub_mc r_lb_mc using sim, replace
quietly {

forvalues i=1/$nmc {

	/*note: for c, y_1, and y_2 equations, "rnormal(0,1)" is residual*/

	replace u=rnormal(0,1)
	replace c=(pi*u)+rnormal(0,1)>1
	replace t_1=(chi*u)+(tau*c)>-0.2
	replace t_2=(gamma*u)+(nu*c)>1
	replace y_1=(delta*t_1)+(psi*u)+rnormal(0,1)
	replace y_2=(delta*t_2)+(psi*u)+(eta*y_1)+rnormal(0,1)
	replace d=y_2-y_1
	
	/*gain-score regression*/
	reg d t_1 t_2 c
	
	/*store regression output*/
	*sibling 1 treatment
	scalar t1_mc=_b[t_1]
	scalar t1_ub_mc=(_b[t_1]+(1.96*_se[t_1])) /*95% CI upper bound*/
	scalar t1_lb_mc=(_b[t_1]-(1.96*_se[t_1])) /*95% CI lower bound*/
	*sibling 2 treatment
	scalar t2_mc=_b[t_2]
	scalar t2_ub_mc=(_b[t_2]+(1.96*_se[t_2])) /*95% CI upper bound*/
	scalar t2_lb_mc=(_b[t_2]-(1.96*_se[t_2])) /*95% CI lower bound*/
	*robustness test result
	scalar r_mc=_b[c] 
	scalar r_ub_mc=(_b[c]+(1.96*_se[c])) /*95% CI upper bound*/
	scalar r_lb_mc=(_b[c]-(1.96*_se[c])) /*95% CI lower bound*/
	
	post `sim' (t1_mc) (t1_ub_mc) (t1_lb_mc) ///
		(t2_mc) (t2_ub_mc) (t2_lb_mc) ///
		(r_mc) (r_ub_mc) (r_lb_mc)
}
}
postclose `sim'

use sim, clear
summarize

**store mean of results, save to append with other simulation results
egen t1=mean(t1_mc)
egen t1_ub=mean(t1_ub_mc)
egen t1_lb=mean(t1_lb_mc)
egen t2=mean(t2_mc)
egen t2_ub=mean(t2_ub_mc)
egen t2_lb=mean(t2_lb_mc)
egen r=mean(r_mc)
egen r_ub=mean(r_ub_mc)
egen r_lb=mean(r_lb_mc)
gen id=5 /*indicates simulation order*/

**generate coverage probability (% of confidence intervals that overlap zero)
gen cov_pre=0
replace cov_pre=1 if r_lb_mc<0 & 0<r_ub_mc
tab cov_pre
egen cov=mean(cov_pre)

keep t1 t1_ub t1_lb t2 t2_ub t2_lb r r_ub r_lb cov id
duplicates drop
duplicates report 

**save data 
save "ph_sim5.dta", replace


/************************************/
*Model 2B, Sim 1.2
*With outcome-to-outcome spillover
*C related to U
*C unrelated to T1
/************************************/

**clear previous data
clear all
set more off

**monte carlo simulation settings 
global nobs=5000 /*observations per run*/
global nmc=1000 /*number of runs*/
set seed 54385 /*seed for random number generator -- can change*/
set obs $nobs /*sets observations within each run*/

**set parameters (causal effects in models)
scalar delta=3 /*t_1 on y_1, t_2 on y_2*/

scalar psi=5 /*u on y_1, u on y_2*/
scalar chi=1 /*u on t_1*/
scalar gamma=2 /*u on t_2*/
scalar pi=0.5 /*u on c*/

scalar eta=0.3 /*y_1 on y_2*/

scalar phi=0.9 /*t_1 on t_2*/
scalar tau=0.7 /*c on t_1*/
scalar nu=0 /*c on t_2*/
scalar lambda=2.5 /*c on y_1, c on y_2*/
scalar omega=2.8 /*c on y_2*/
scalar theta=1.2 /*t_1 on y_2*/
scalar kappa=0.6 /*t_2 on y_1*/
scalar mu=0.4 /*y_1 on t_2*/

**generate variables

gen u=. /*unobserved fixed effect*/
gen c=. /*observed family-level characteristic*/ 
gen t_1=. /*sibling 1's treatment*/
gen t_2=. /*sibling 2's treatment*/
gen y_1=. /*sibling 1's outcome*/
gen y_2=. /*sibling 2's outcome*/
gen d=. /*gain-score (y_2-y_1)*/

**simulation 

tempname sim
/*
for t1, t2, and c:
X_mc is the regression coefficient
X_ub_mc is the 95% confidence interval upper bound
X_lb_mc is the 95% confidence interval lower bound
*/
postfile `sim' t1_mc t1_ub_mc t1_lb_mc ///
	t2_mc t2_ub_mc t2_lb_mc ///
	r_mc r_ub_mc r_lb_mc using sim, replace
quietly {

forvalues i=1/$nmc {

	/*note: for c, y_1, and y_2 equations, "rnormal(0,1)" is residual*/

	replace u=rnormal(0,1)
	replace c=(pi*u)+rnormal(0,1)>1
	replace t_1=(chi*u)+(tau*c)>-0.2
	replace t_2=(gamma*u)+(nu*c)>1
	replace y_1=(delta*t_1)+(psi*u)+rnormal(0,1)
	replace y_2=(delta*t_2)+(psi*u)+(eta*y_1)+rnormal(0,1)
	replace d=y_2-y_1
	
	/*gain-score regression*/
	reg d t_1 t_2 c
	
	/*store regression output*/
	*sibling 1 treatment
	scalar t1_mc=_b[t_1]
	scalar t1_ub_mc=(_b[t_1]+(1.96*_se[t_1])) /*95% CI upper bound*/
	scalar t1_lb_mc=(_b[t_1]-(1.96*_se[t_1])) /*95% CI lower bound*/
	*sibling 2 treatment
	scalar t2_mc=_b[t_2]
	scalar t2_ub_mc=(_b[t_2]+(1.96*_se[t_2])) /*95% CI upper bound*/
	scalar t2_lb_mc=(_b[t_2]-(1.96*_se[t_2])) /*95% CI lower bound*/
	*robustness test result
	scalar r_mc=_b[c] 
	scalar r_ub_mc=(_b[c]+(1.96*_se[c])) /*95% CI upper bound*/
	scalar r_lb_mc=(_b[c]-(1.96*_se[c])) /*95% CI lower bound*/
	
	post `sim' (t1_mc) (t1_ub_mc) (t1_lb_mc) ///
		(t2_mc) (t2_ub_mc) (t2_lb_mc) ///
		(r_mc) (r_ub_mc) (r_lb_mc)
}
}
postclose `sim'

use sim, clear
summarize

**store mean of results, save to append with other simulation results
egen t1=mean(t1_mc)
egen t1_ub=mean(t1_ub_mc)
egen t1_lb=mean(t1_lb_mc)
egen t2=mean(t2_mc)
egen t2_ub=mean(t2_ub_mc)
egen t2_lb=mean(t2_lb_mc)
egen r=mean(r_mc)
egen r_ub=mean(r_ub_mc)
egen r_lb=mean(r_lb_mc)
gen id=6 /*indicates simulation order*/

**generate coverage probability (% of confidence intervals that overlap zero)
gen cov_pre=0
replace cov_pre=1 if r_lb_mc<0 & 0<r_ub_mc
tab cov_pre
egen cov=mean(cov_pre)

keep t1 t1_ub t1_lb t2 t2_ub t2_lb r r_ub r_lb cov id
duplicates drop
duplicates report 

**save data 
save "ph_sim6.dta", replace


/************************************/
*Model 2B, Sim 2.1
*No outcome-to-outcome spillover
*C unrelated to U
*C unrelated to T2
/************************************/

**clear previous data
clear all
set more off

**monte carlo simulation settings 
global nobs=5000 /*observations per run*/
global nmc=1000 /*number of runs*/
set seed 54385 /*seed for random number generator -- can change*/
set obs $nobs /*sets observations within each run*/

**set parameters (causal effects in models)
scalar delta=3 /*t_1 on y_1, t_2 on y_2*/

scalar psi=5 /*u on y_1, u on y_2*/
scalar chi=1 /*u on t_1*/
scalar gamma=2 /*u on t_2*/
scalar pi=0 /*u on c*/

scalar eta=0 /*y_1 on y_2*/

scalar phi=0.9 /*t_1 on t_2*/
scalar tau=0.7 /*c on t_1*/
scalar nu=0 /*c on t_2*/
scalar lambda=2.5 /*c on y_1, c on y_2*/
scalar omega=2.8 /*c on y_2*/
scalar theta=1.2 /*t_1 on y_2*/
scalar kappa=0.6 /*t_2 on y_1*/
scalar mu=0.4 /*y_1 on t_2*/

**generate variables

gen u=. /*unobserved fixed effect*/
gen c=. /*observed family-level characteristic*/ 
gen t_1=. /*sibling 1's treatment*/
gen t_2=. /*sibling 2's treatment*/
gen y_1=. /*sibling 1's outcome*/
gen y_2=. /*sibling 2's outcome*/
gen d=. /*gain-score (y_2-y_1)*/

**simulation 

tempname sim
/*
for t1, t2, and c:
X_mc is the regression coefficient
X_ub_mc is the 95% confidence interval upper bound
X_lb_mc is the 95% confidence interval lower bound
*/
postfile `sim' t1_mc t1_ub_mc t1_lb_mc ///
	t2_mc t2_ub_mc t2_lb_mc ///
	r_mc r_ub_mc r_lb_mc using sim, replace
quietly {

forvalues i=1/$nmc {

	/*note: for c, y_1, and y_2 equations, "rnormal(0,1)" is residual*/

	replace u=rnormal(0,1)
	replace c=(pi*u)+rnormal(0,1)>1
	replace t_1=(chi*u)+(tau*c)>-0.2
	replace t_2=(gamma*u)+(nu*c)>1
	replace y_1=(delta*t_1)+(psi*u)+rnormal(0,1)
	replace y_2=(delta*t_2)+(psi*u)+(eta*y_1)+rnormal(0,1)
	replace d=y_2-y_1
	
	/*gain-score regression*/
	reg d t_1 t_2 c
	
	/*store regression output*/
	*sibling 1 treatment
	scalar t1_mc=_b[t_1]
	scalar t1_ub_mc=(_b[t_1]+(1.96*_se[t_1])) /*95% CI upper bound*/
	scalar t1_lb_mc=(_b[t_1]-(1.96*_se[t_1])) /*95% CI lower bound*/
	*sibling 2 treatment
	scalar t2_mc=_b[t_2]
	scalar t2_ub_mc=(_b[t_2]+(1.96*_se[t_2])) /*95% CI upper bound*/
	scalar t2_lb_mc=(_b[t_2]-(1.96*_se[t_2])) /*95% CI lower bound*/
	*robustness test result
	scalar r_mc=_b[c] 
	scalar r_ub_mc=(_b[c]+(1.96*_se[c])) /*95% CI upper bound*/
	scalar r_lb_mc=(_b[c]-(1.96*_se[c])) /*95% CI lower bound*/
	
	post `sim' (t1_mc) (t1_ub_mc) (t1_lb_mc) ///
		(t2_mc) (t2_ub_mc) (t2_lb_mc) ///
		(r_mc) (r_ub_mc) (r_lb_mc)
}
}
postclose `sim'

use sim, clear
summarize

**store mean of results, save to append with other simulation results
egen t1=mean(t1_mc)
egen t1_ub=mean(t1_ub_mc)
egen t1_lb=mean(t1_lb_mc)
egen t2=mean(t2_mc)
egen t2_ub=mean(t2_ub_mc)
egen t2_lb=mean(t2_lb_mc)
egen r=mean(r_mc)
egen r_ub=mean(r_ub_mc)
egen r_lb=mean(r_lb_mc)
gen id=7 /*indicates simulation order*/

**generate coverage probability (% of confidence intervals that overlap zero)
gen cov_pre=0
replace cov_pre=1 if r_lb_mc<0 & 0<r_ub_mc
tab cov_pre
egen cov=mean(cov_pre)

keep t1 t1_ub t1_lb t2 t2_ub t2_lb r r_ub r_lb cov id
duplicates drop
duplicates report 

**save data 
save "ph_sim7.dta", replace


/************************************/
*Model 2B, Sim 2.2
*With outcome-to-outcome spillover
*C unrelated to U
*C unrelated to T2
/************************************/

**clear previous data
clear all
set more off

**monte carlo simulation settings 
global nobs=5000 /*observations per run*/
global nmc=1000 /*number of runs*/
set seed 54385 /*seed for random number generator -- can change*/
set obs $nobs /*sets observations within each run*/

**set parameters (causal effects in models)
scalar delta=3 /*t_1 on y_1, t_2 on y_2*/

scalar psi=5 /*u on y_1, u on y_2*/
scalar chi=1 /*u on t_1*/
scalar gamma=2 /*u on t_2*/
scalar pi=0 /*u on c*/

scalar eta=0.3 /*y_1 on y_2*/

scalar phi=0.9 /*t_1 on t_2*/
scalar tau=0.7 /*c on t_1*/
scalar nu=0 /*c on t_2*/
scalar lambda=2.5 /*c on y_1, c on y_2*/
scalar omega=2.8 /*c on y_2*/
scalar theta=1.2 /*t_1 on y_2*/
scalar kappa=0.6 /*t_2 on y_1*/
scalar mu=0.4 /*y_1 on t_2*/

**generate variables

gen u=. /*unobserved fixed effect*/
gen c=. /*observed family-level characteristic*/ 
gen t_1=. /*sibling 1's treatment*/
gen t_2=. /*sibling 2's treatment*/
gen y_1=. /*sibling 1's outcome*/
gen y_2=. /*sibling 2's outcome*/
gen d=. /*gain-score (y_2-y_1)*/

**simulation 

tempname sim
/*
for t1, t2, and c:
X_mc is the regression coefficient
X_ub_mc is the 95% confidence interval upper bound
X_lb_mc is the 95% confidence interval lower bound
*/
postfile `sim' t1_mc t1_ub_mc t1_lb_mc ///
	t2_mc t2_ub_mc t2_lb_mc ///
	r_mc r_ub_mc r_lb_mc using sim, replace
quietly {

forvalues i=1/$nmc {

	/*note: for c, y_1, and y_2 equations, "rnormal(0,1)" is residual*/

	replace u=rnormal(0,1)
	replace c=(pi*u)+rnormal(0,1)>1
	replace t_1=(chi*u)+(tau*c)>-0.2
	replace t_2=(gamma*u)+(nu*c)>1
	replace y_1=(delta*t_1)+(psi*u)+rnormal(0,1)
	replace y_2=(delta*t_2)+(psi*u)+(eta*y_1)+rnormal(0,1)
	replace d=y_2-y_1
	
	/*gain-score regression*/
	reg d t_1 t_2 c
	
	/*store regression output*/
	*sibling 1 treatment
	scalar t1_mc=_b[t_1]
	scalar t1_ub_mc=(_b[t_1]+(1.96*_se[t_1])) /*95% CI upper bound*/
	scalar t1_lb_mc=(_b[t_1]-(1.96*_se[t_1])) /*95% CI lower bound*/
	*sibling 2 treatment
	scalar t2_mc=_b[t_2]
	scalar t2_ub_mc=(_b[t_2]+(1.96*_se[t_2])) /*95% CI upper bound*/
	scalar t2_lb_mc=(_b[t_2]-(1.96*_se[t_2])) /*95% CI lower bound*/
	*robustness test result
	scalar r_mc=_b[c] 
	scalar r_ub_mc=(_b[c]+(1.96*_se[c])) /*95% CI upper bound*/
	scalar r_lb_mc=(_b[c]-(1.96*_se[c])) /*95% CI lower bound*/
	
	post `sim' (t1_mc) (t1_ub_mc) (t1_lb_mc) ///
		(t2_mc) (t2_ub_mc) (t2_lb_mc) ///
		(r_mc) (r_ub_mc) (r_lb_mc)
}
}
postclose `sim'

use sim, clear
summarize

**store mean of results, save to append with other simulation results
egen t1=mean(t1_mc)
egen t1_ub=mean(t1_ub_mc)
egen t1_lb=mean(t1_lb_mc)
egen t2=mean(t2_mc)
egen t2_ub=mean(t2_ub_mc)
egen t2_lb=mean(t2_lb_mc)
egen r=mean(r_mc)
egen r_ub=mean(r_ub_mc)
egen r_lb=mean(r_lb_mc)
gen id=8 /*indicates simulation order*/

**generate coverage probability (% of confidence intervals that overlap zero)
gen cov_pre=0
replace cov_pre=1 if r_lb_mc<0 & 0<r_ub_mc
tab cov_pre
egen cov=mean(cov_pre)

keep t1 t1_ub t1_lb t2 t2_ub t2_lb r r_ub r_lb cov id
duplicates drop
duplicates report 

**save data 
save "ph_sim8.dta", replace


/************************************/
*Compile post-hoc simulation results
/************************************/

clear
set more off

**open first simulation dataset 
use "ph_sim1.dta"

**append subsequent simulation datasets
append using "ph_sim2.dta"
append using "ph_sim3.dta"
append using "ph_sim4.dta"
append using "ph_sim5.dta"
append using "ph_sim6.dta"
append using "ph_sim7.dta"
append using "ph_sim8.dta"

**label id (assign to models)
label define model_lbl 1 "Sim 1.1, C->T2" ///
	2 "Sim 1.2, C->T2" ///
	3 "Sim 2.1, C->T2" ///
	4 "Sim 2.2, C->T2" ///
	5 "Sim 1.1, C->T1" ///
	6 "Sim 1.2, C->T1" ///
	7 "Sim 2.1, C->T1" ///
	8 "Sim 2.2, C->T1" ///

**tabulate results
***sim 1.1
list id t1 t1_lb t1_ub t2 t2_lb t2_ub r r_lb r_ub cov if id==1 | id==5
***sim 1.2
list id t1 t1_lb t1_ub t2 t2_lb t2_ub r r_lb r_ub cov if id==2 | id==6
***sim 2.1
list id t1 t1_lb t1_ub t2 t2_lb t2_ub r r_lb r_ub cov if id==3 | id==7
***sim 2.2
list id t1 t1_lb t1_ub t2 t2_lb t2_ub r r_lb r_ub cov if id==4 | id==8 
\end{lstlisting}
\end{singlespace}


\begin{thebibliography}{9}

\bibitem{frisell2012}Frisell T, Öberg S, Kuja-Halkola R, Sjölander A. Sibling comparison designs: bias from non-shared confounders and measurement error. \textit{Epidemiol}. 2012;23:713-720.

\bibitem{khashan2014}Khashan AS, Kenny LC, Lundholm C, Kearney PM, Gong T, Almqvist C. Mode of obstetrical delivery and type 1 diabetes: a sibling design study. \textit{Pediatrics}. 2014;134:e806-813.

\bibitem{khashan2015}Khashan AS, Kenny LC, Lundholm C, et al. Gestational age and birth weight and the risk of childhood type 1 diabetes: a population-based cohort and sibling design study. \textit{Diabetes Care}. 2015;38:2308-2315.

\bibitem{hvolgaard2016}Hvolgaard Mikkelsen S, Olsen J, Bech BH, Obel C. Parental age and attention-deficit/hyperactivity disorder (adhd). \textit{Int J Epidemiol}. 2016;46:409-420.

\bibitem{sjolander2016}Sjölander A, Frisell T, Kuja-Halkola R, Öberg S, Zetterqvist J. Carryover effects in sibling comparison designs. \textit{Epidemiol}. 2016;27:852-858.

\bibitem{hanley2017}Hanley GE, Hutcheon JA, Kinniburgh BA, et al.  Interpregnancy interval and adverse pregnancy outcomes: an analysis of successive pregnancies. \textit{Obstet Gynecol}. 2017;129:408-415.

\bibitem{axelsson2019}Axelsson PB, Clausen TD, Petersen AH, et al. Relation between infant microbiota and autism?: results from a national cohort sibling design study. \textit{Epidemiol}. 2019;30:52-60.

\bibitem{mallinson2020}Mallinson DC, Larson A, Berger LM, Grodsky E, Ehrenthal DB. Estimating the effect of Prenatal Care Coordination in Wisconsin: a sibling fixed effects analysis. \textit{Health Serv Res}. 2020;55:82-93.

\bibitem{petersen2020}Petersen AH, Lange T. What is the causal interpretation of sibling comparison designs? \textit{Epidemiol}. 2020;31:75-81.

\bibitem{frisell2021}Sibling-comparison designs, are they worth the effort? \textit{Am J Epidemiol}. 2021;190:738-741.

\bibitem{mallinson2021}Mallinson DC, Elwert FE. Estimating sibling spillover effects with unobserved confounding using gain-scores. arXiv, doi:2102.11150v3, 6 May 2021, preprint: not peer reviewed. 

\bibitem{feinberg2012}Feinberg ME, Solmeyer AR, McHale SM. The third rail of family systems: sibling relationships, mental and behavioral health, and preventive intervention in childhood and adolescence. \textit{Clin Child Fam Psychol Rev}. 2012;15:43–57.

\bibitem{deneve2017}De Neve JW, Kawachi I. Spillovers between siblings and from offspring to parents are understudied: a review and future directions for research. \textit{Soc Sci Med}. 2017;183:56-61.

\bibitem{lipsitch2010}Lipsitch M, Tchetgen Tchetgen E, Cohen T. Negative controls: a tool for detecting confounding and bias in observational studies. \textit{Epidemiol}. 2010;21:383-388.

\bibitem{cunninghambook}Cunningham S. Causal Inference: The Mixtape. New Haven: Yale University Press, 2021.

\bibitem{kim2019}Kim Y, Steiner PM. Gain scores revisited: a graphical models perspective. \textit{Sociol Methods Res}. 2021;50:1353-1375.

\bibitem{kim2021}Kim Y, Steiner PM. Causal graphical views of fixed effects and random effects models. \textit{Br J Math Stat Psychol}. 2021;74:165-183.

\bibitem{gunasekara2014}Gunasekara FI, Richardson K, Carter K, Blakely T. Fixed effects analysis of repeated measures data. \textit{Int J Epidemiol}. 2014;43:264-269.

\bibitem{imai2019}Imai K, Kim IS. When should we use unit fixed effects regression models for causal inference with longitudinal data? \textit{Am J Pol Sci}. 2019;63:467-490.

\bibitem{qin2010}Qin C, Gould JB. Maternal nativity status and birth outcomes in Asian immigrants. \textit{J Immigrant Minority Health}. 2010;12:798-805. 

\bibitem{creanga2012}Creanga AA, Berg CJ, Syverson C, Seed K, Bruce FC, Callaghan WM. Race, ethnicity, and nativity differentials in pregnancy-related mortality in the United States. \textit{Obstet Gynecol}. 2012;120:361-268.

\bibitem{elwert2014}Elwert F, Winship C. Endogenous selection bias: the problem of conditioning on a collider variable. \textit{Annu Rev Soc}. 2014;40:31-53. 

\bibitem{vwbook}VanderWeele TJ. Explanation in Causal Inference: Methods for Mediation and Interaction. New York: Oxford University Press, 2015.

\bibitem{adkins2012}Adkins LC, Gade MN. Monte Carlo experiments using Stata: a primer with examples. \textit{Adv Econ}. 2012;30:429-77.

\bibitem{statasoft}StataCorp. Stata Statistical Software: Release 16. College Station, TX: StataCorp LLC; 2019.

\bibitem{nchs}National Center for Health Statistics. Revisions of the U.S. Standard Certificates and Reports. https://www.cdc.gov/nchs/nvss/revisions-of-the-us-standard-certificates-and-reports.htm. Accessed September 3, 2021.

\bibitem{pearl2013}Pearl J. Linear models: a useful "microscope" for causal analysis. \textit{J Causal Inference}. 2013;1:155-170.

\bibitem{tennant2021}Tennant PWG, Murray EJ, Arnold KF, et al. Use of directed acyclic graphs (DAGs) to identify confounders in applied health research: review and recommendations. \textit{Int J Epidemiol}. 2021;50:620-632.

\bibitem{sjolander2017}Sjölander A, Zetterqvist J. Confounders, mediators, or colliders. \textit{Epidemiol}. 2017;28:540-547.

\end{thebibliography}
\end{document}